\documentclass[aps,twocolumn,superscriptaddress,showpacs,showkeys, notitlepage, longbibliography]{revtex4-1}

\usepackage[bookmarks=false,colorlinks,citecolor=blue,linkcolor=blue,urlcolor=blue]{hyperref}
\usepackage{graphicx}
\usepackage{float}

\usepackage{tikz,pdfpages}
\makeatletter
\AtBeginDocument{\let\LS@rot\@undefined}
\makeatother

\usepackage{dcolumn}
\usepackage{bm}
\usepackage{float}
\usepackage{amsmath}
\usepackage{bm}
\usepackage[normalem]{ulem}
\usepackage{xcolor}
\begin{document}
	
	\title{Dirac-vortex topological photonic crystal fibre }
	
	\author{Hao Lin}
	\affiliation{Institute of Physics, Chinese Academy of Sciences/Beijing National Laboratory for Condensed Matter Physics, Beijing 100190}
	\affiliation{School of Physical Sciences, University of Chinese Academy of Sciences, Beijing 100049, China}
	\author{Ling Lu}
	\email{linglu@iphy.ac.cn}
	\affiliation{Institute of Physics, Chinese Academy of Sciences/Beijing National Laboratory for Condensed Matter Physics, Beijing 100190}
	\affiliation{Songshan Lake Materials Laboratory, Dongguan, Guangdong 523808, China}

	\begin{abstract}
	The success of photonic crystal fibres relies largely on the endless variety of two-dimensional photonic crystals in the cross-section. Here, we propose a topological bandgap fibre whose bandgaps along in-plane directions are opened by generalized $\rm{Kekul\acute{e}}$ modulation of a Dirac lattice with a vortex phase. Then, the existence of mid-gap defect modes is guaranteed to guide light at the core of this Dirac-vortex fibre, where the number of guiding modes equals the winding number of the spatial vortex. The single-vortex design provides a single-polarization single mode for a bandwidth as large as one octave.	
	\end{abstract}

	\maketitle

\section{Introduction}
Topological photonics~\cite{lu2014topological,khanikaev2017two,ozawa2019topological}, initiated with the idea of robust waveguiding, is inspiring novel fibre concepts, such as a one-way fibre inside a magnetic three-dimensional photonic crystal~\cite{lu2018topological} and a Bragg fibre with nontrivial edge modes~\cite{pilozzi2020topological}. In this article, we introduce the topological photonic crystal fibre~(PCF) whose invariant cross-section resembles the recent Dirac-vortex topological cavity~\cite{gao2020dirac} in two-dimensional photonic crystals. Such a topological bound state can be traced back to the Jackiw-Rossi zero mode in the 2D Dirac equation~\cite{jackiw1981zero}, and has been realized in honeycomb lattices~\cite{hou2007electron,iadecola2016non} in a couple of systems~\cite{menssen2020photonic,noh2020braiding,gao2019majorana,chen2019mechanical}. This Dirac-vortex silica fibre can support an \emph{arbitrary} number of nearly degenerate guiding modes by varying the winding number~($w$) of the spatial vortex. When $w=\pm1$, the fibre can support a single-polarization single mode~(SPSM) with a large bandwidth.

The SPSM fibre supports truly one mode, while traditional single-mode fibres and polarization-maintaining fibres both support two polarizations, either degenerate or nondegenerate, respectively. Such fibre birefringence (dual polarization) broadens the optical pulses being transmitted, known as polarization-mode dispersion. To solve this limitation, SPSM fibres have been designed to separate the degenerate cutoff frequencies by lowering the symmetry of the fibre cross-section, which can be achieved by structural asymmetry or nonuniform stress. This dominant asymmetric approach is applied mostly to the lowest-frequency index-guided fundamental modes~(polarization-degenerate)~\cite{okoshi1980single,eickhoff1982stress,simpson1983single,kubota2004absolutely,folkenberg2005broadband,lee2008design}, but also works for the nonfundamental degenerate modes inside a bandgap~\cite{ferrando2001single,eguchi2012single,szpulak2007single}.

\begin{figure}[h]
	\centering
	\includegraphics[width=0.5\textwidth]{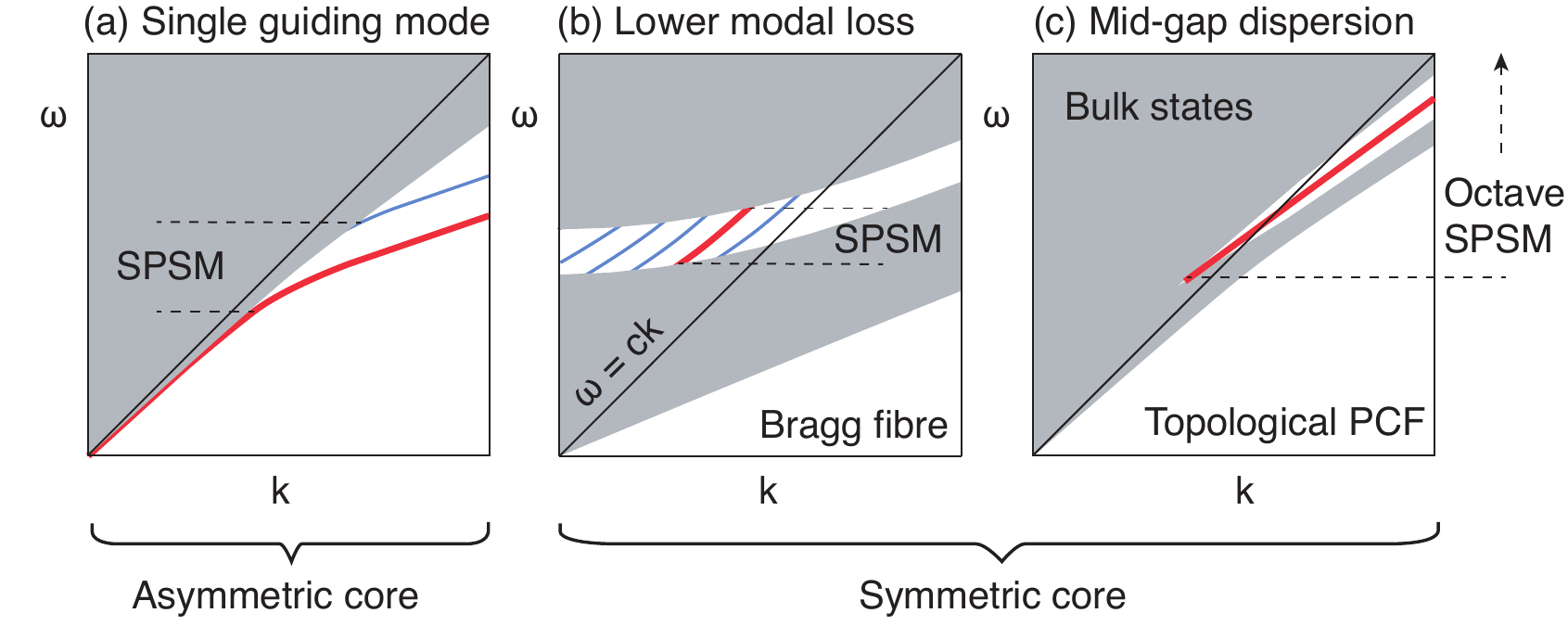}
	\caption{
	Asymmetric~(a) and symmetric~(b, c) approaches to an SPSM.
	(a) Dominant approach to split the degenerate fundamental modes by structural asymmetry.
	(b)	Proposed Bragg fibre design to guide light by a singly degenerate mode with lower loss than other modes.
	(c) The topological PCF provides singlet mid-gap dispersion for a broadband SPSM.
}
	\label{fig::intro}
	\end{figure}

The SPSM bandwidth is very limited due to the amount of asymmetry that can be applied; the best experimental value is approximately 30\%~(frequency span over central frequency), which is achieved in a stressed PCF~\cite{folkenberg2005broadband}.
(In integrated waveguides, an SPSM bandwidth over one octave~[66.67\%] has been demonstrated~\cite{chiles2016demonstration}.)
For example, in Fig.~\ref{fig::intro}(a), the SPSM bandwidth is bounded from above by the cutoff frequency of the other polarization~(blue) and bounded from below by the confinement loss, even if this guiding mode~(red line) has no cutoff frequency. 
However, there exists a symmetric approach to an SPSM by operating at a singly degenerate mode inside a bandgap so that the fibre cross-section can remain highly symmetric. Shown in Fig.~\ref{fig::intro}(b), this symmetric approach has thus far only been proposed theoretically in a hollow-core Bragg fibre
~\cite{bassett2002elimination,argyros2004microstructured}, in which many other guided modes (blue line) can be made much lossier than the TE$_{01}$ mode (red line). The SPSM bandwidth of this design is even more limited.

The Dirac-vortex fibre is an ideal design for an ultrabroadband SPSM by ensuring singlet mid-gap dispersion inside the bandgap (the symmetric approach), as illustrated in Fig.~\ref{fig::intro}(c). Without the topological mechanism of the Dirac vortex, it is generally difficult to stabilize a defect mode at the middle of the bandgap for every wave vector.

In the rest of the article, we first show how to gap the nodal line without leaving residual Weyl points. Such a supercell ${\rm{Kekul\acute{e}}}$ modulation~\cite{hou2007electron,gao2020dirac} is generalized to continuous $2\pi$ phase angles that are used to construct a vortex gap around the fibre core that can confine any number of fibre modes. To ease the fabrication, a simplified design of only four capillary silica tubes is introduced. Finally, we enlarge the vortex size to eliminate the index-guided modes and achieve an octave-spanning SPSM.

	\begin{figure}[b]
	\centering
	\includegraphics[width=0.45\textwidth]{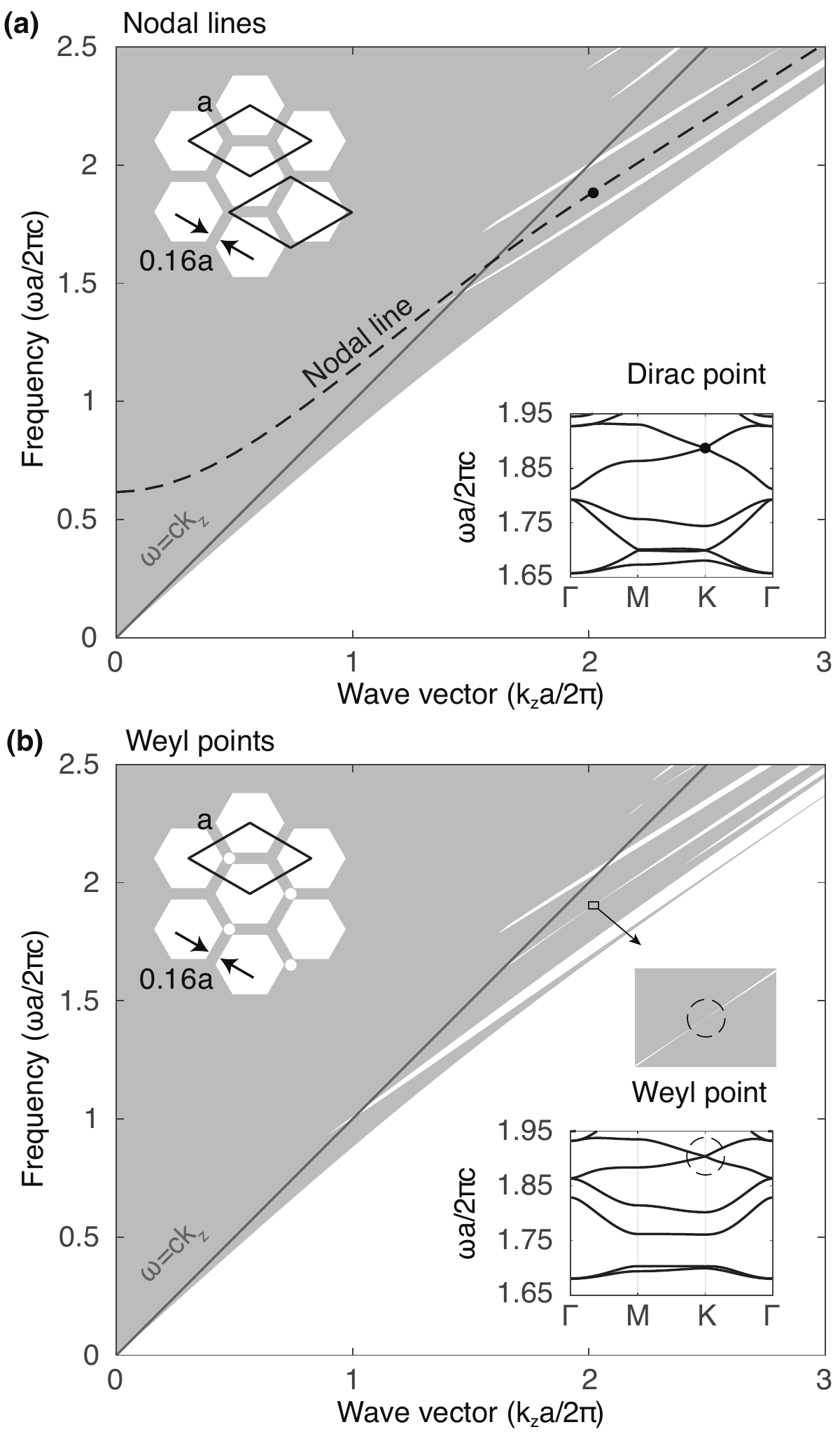}
	\caption{Band diagram of silica~($\varepsilon=2.1$) photonic crystal lattices uniform in the out-of-plane direction~($z$).
	(a) Projected band diagram of the triangular photonic crystal, in which the nodal line degeneracy is highlighted.
	(b) An extra air hole in the primitive cell breaks the inversion  symmetry, and the nodal line is lifted into Weyl points. Insets: cross-section structures and in-plane band structures at $ k_{z}a/2\pi=2.02$.
Two different primitive-cell choices are drawn in (a).}
	\label{fig::nodalline}
	\end{figure}
	
\begin{figure*}[th!]
		\centering
		\includegraphics[width=\textwidth]{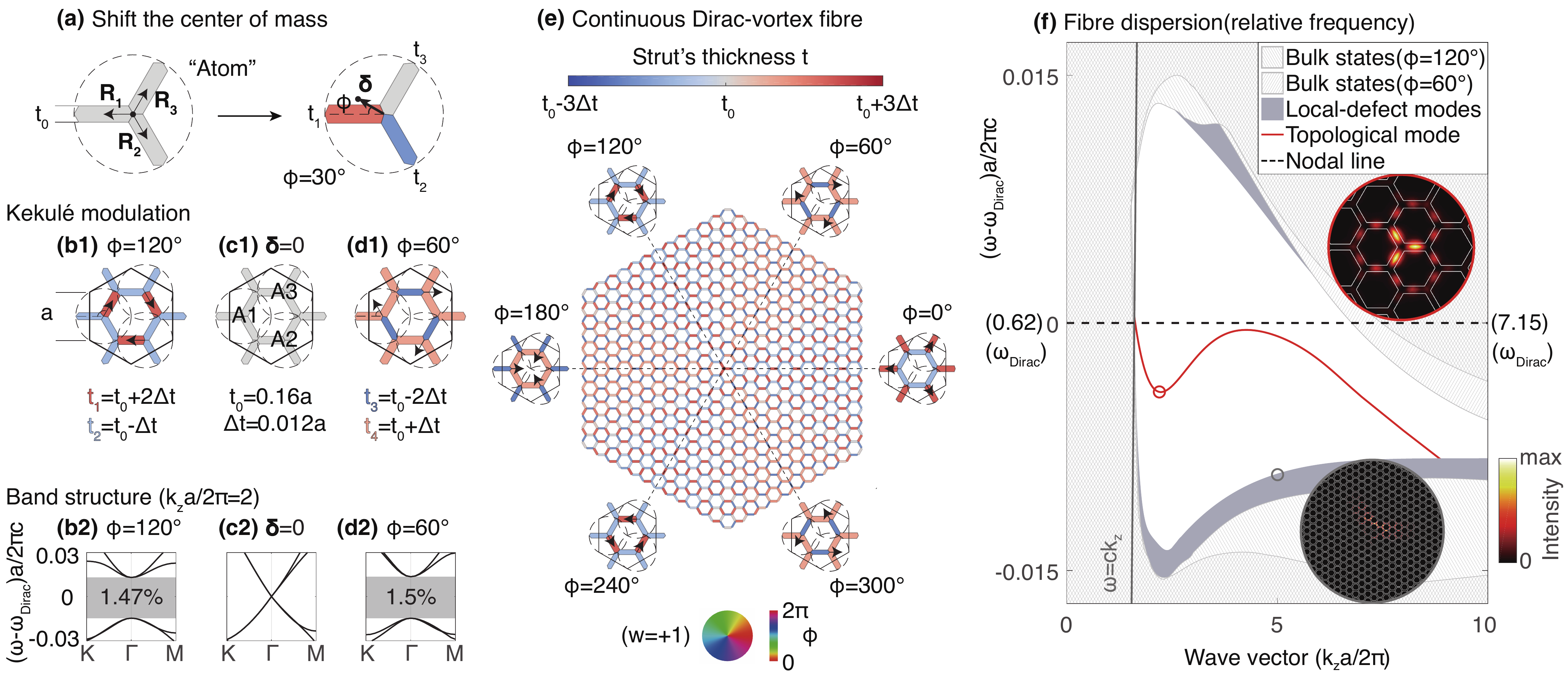}
		\caption{Dirac-vortex fibre obtained by continuous Kekul$\rm\acute{e}$ modulations.
		(a) Example of how an ``atom'' can be shifted in any direction~($\arg[\boldsymbol{\delta}] =\phi$) with finite amplitude~($|\boldsymbol{\delta}|$) by changing the widths of the three struts.
(b1), (c1), (d1) Supercell examples of three coordinated atoms~(A1, A2, A3) with $\left| \boldsymbol{\delta} \right|=\sqrt{3}a/80$, $\phi=120^\circ$; $|\boldsymbol{\delta}|=0$; and $\left| \boldsymbol{\delta} \right|=\sqrt{3}a/80$, $\phi=60^\circ$, respectively. $\Delta t=2\sqrt{3} |\boldsymbol{\delta} | t_0/a = 0.012a$.
The corresponding band structures are plotted in (b2), (c2), (d2), respectively.
(e) Structure of a continuous Dirac-vortex PCF, in which every strut is coloured according to its width.
(f) Band diagram of the fibre plotted in reference to the frequency of the original nodal line~(central dashed line). The inset shows the intensity patterns~($\hat{\boldsymbol{ z}}\cdot \rm Re[\boldsymbol{ E}^{*} \times \boldsymbol{ H}]$) of the topological mode and one local-defect mode. The single-polarization topological mode~(red line) spans over two octaves.
}
		\label{fig::continuous}
	\end{figure*}
	
		\begin{figure*}[t]
		\centering
		\includegraphics[width=1\textwidth]{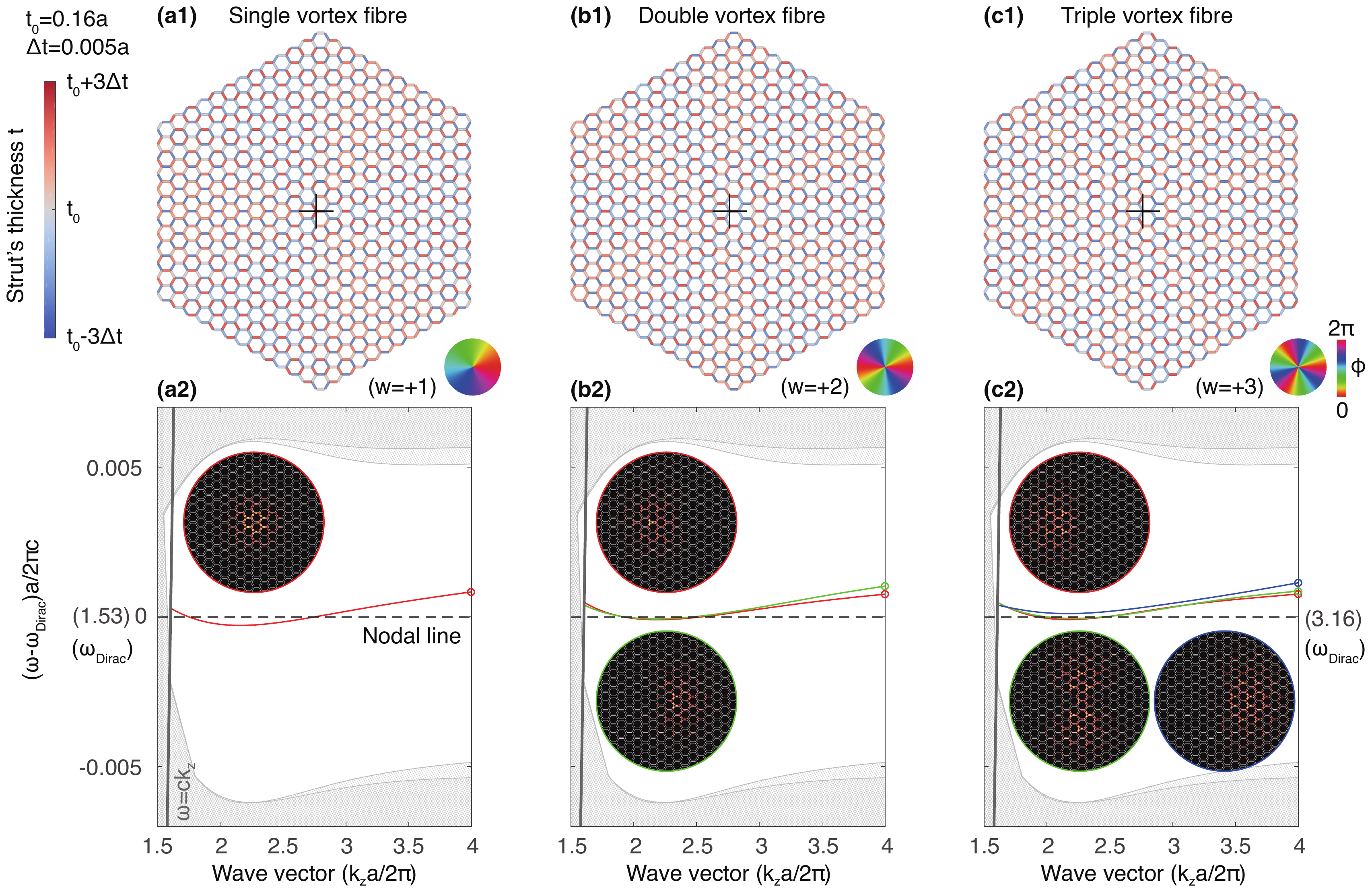}
		\caption{Continuous Dirac-vortex fibres with winding number $w=+1,+2,+3$.
		(a1), (b1), and (c1) are the fibre structures. The colour wheels show the corresponding phases of the generalized $\rm{Kekul\acute{e}}$ modulation. (a2), (b2), and (c2) are the corresponding band diagrams, and the insets show the mode profiles~($\hat{\boldsymbol{ z}}\cdot \rm Re[\boldsymbol{ E}^{*} \times \boldsymbol{ H}]$) of modes at $k_{z}a/2\pi=4$. }
	\label{fig::multimode}
	\end{figure*}	

\section{Results}	
\subsection{ Nodal lines and Weyl points in a PCF}
We start with the most common PCF structure~\cite{knight2003photonic,russell2006photonic}, a silica photonic crystal with a triangular lattice of air holes, shown in Fig.~\ref{fig::nodalline}(a). There are two nodal lines of 2D Dirac points~\cite{xie2015fiber,biswas2016dirac}, at the $\pm K$ points in the Brillouin zone. The Dirac points are frequency isolated in the in-plane 2D band structure for $k_{z}a/2\pi>1.6$.
Each nodal line gaps into Weyl points~\cite{lu2013weyl}, if we break the inversion symmetry by adding an extra small air hole in the primitive cell, as shown in Fig.~\ref{fig::nodalline}(b).
Since this PCF structure has mirror symmetry along $z$, the Weyl surface state does not have handedness propagating along the $z$ axis, unlike that observed in coupled spiral waveguide arrays~\cite{rechtsman2013photonic,noh2017experimental} containing similar type-II Weyl points. To gap the entire nodal line, without leaving Weyl degeneracies, we apply supercell modulations.

\subsection{Generalized Kekul$\mathbf{\acute{e}}$ modulation}

As shown in Fig.~\ref{fig::continuous}(a-d), we perturb the nodal-line lattice with the generalized $\rm{Kekul\acute{e}}$ pattern discussed in Ref.~\cite{gao2020dirac}. The idea is to couple the two nodal lines~(of Dirac points) together in an enlarged supercell and annihilate them into a bandgap through supercell modulation. Since each supercell has three primitive cells, we label each primitive cell~[Fig.~\ref{fig::nodalline}(a) inset] as an artificial ``atom'' consisting of three struts~[Fig.~\ref{fig::continuous}(a)].
In practice, we shift the three ``atoms'' in the supercell with identical amplitude and constant phase difference~(120$^\circ$) between each two ``atoms''.
To move each ``atom'', we can shift its centre of mass~$\boldsymbol{\delta}(\phi)$, in any direction $\phi$, by adjusting the thickness of the three struts~($t_1$, $t_2$, $t_3$) without changing the total mass of the ``atom''~($\propto 3t_0$). Therefore, the strut thickness for any modulation vector $\boldsymbol{\delta}(\phi)$ can be determined by the following equation:
	\begin{equation} 
	\sum_{i=1}^3t_i \boldsymbol {R}_i =  \boldsymbol{\delta} (\phi) \sum_{i=1}^3t_i= \boldsymbol{\delta} (\phi) 3t_0
	\end{equation}
, where $\boldsymbol{R}_i$~($|\boldsymbol{R}_i|=\sqrt{3}/6a$) is the position vector of the centre of mass of each strut. Examples of the lattices before and after the $\rm{Kekul\acute{e}}$ modulations are drawn in Fig.~\ref{fig::continuous}(b1-d1), and their corresponding band structures are plotted in (b2-d2).

After the generalized modulation~($|\boldsymbol{\delta}(\phi)|\neq0$),
the entire nodal lines gap out for arbitrary $\phi\in(0,2\pi)$, and the bandgaps share a common frequency range, which is eventually determined by the band edges of the two most symmetric supercells of $\phi$=60$^\circ$ and 120$^\circ$.
The lattices of the rest of the modulation phases can be understood as interpolations between the two. Note that the modulation phase $\phi$ of each supercell is labelled by the shift angle of ``atom'' A1 in Fig.~\ref{fig::continuous}(c1).

\subsection{Continuous modulation}
As shown in Fig.~\ref{fig::continuous}(e), we arrange the series of $\rm{Kekul\acute{e}}$-modulated 2D lattices angularly around a chosen core point in the fibre cross-section as a function of their modulation phase $\phi$.As a result, a single defect mode appears in the middle of the bandgap and is spatially localized at the fibre core, as shown in Fig.~\ref{fig::continuous}(f). This result is expected from the Dirac-vortex cavity result~\cite{gao2020dirac} because a PCF can be regarded as a 2D cavity for every propagation constant~($k_z$).
Since the fibre is $C_{3v}$ symmetric, one-sixth of the structure is sufficient to study this topological mode.

    The large momentum~[$k_z a/(2\pi)>5$], or short wavelength, behaviour of the fibre band diagram in Fig.~\ref{fig::continuous}(f) is worth discussion. First, the common bandgap frequency is lower than that of the nodal line~(central dashed line in the figure) at short wavelengths because the short-wavelength modes can localize preferentially at struts that are thicker than $t_0$~(rather than at those that are thinner) in the modulated lattice, resulting in lowering of the overall band frequencies. Second, the topological mode is very sensitive to the central three struts, where the modal intensity peaks. In this case, the three central red struts are the thickest, and consequently, the fibre mode cuts off, in the short-wavelength region, at a frequency lower than the bandgap. Third, although we “continuously" vary the modulation angle $\phi$ in the surrounding lattice, the discretization is still limited to a single strut in our design. A consequence of the strut-discrete vortex is that a set of extra modes is trapped very close to the spectral edges of the bandgap. We call them local-defect modes, shaded in grey colour in the figure. These modes tend to arise at short wavelengths and are more prominent for large modulation amplitudes~(large bandgaps), as shown in the example in Fig.~\ref{fig::continuous}(f). Since these local-defect modes are located near the bulk band edges with an extended mode profile, their confinement losses are at least one order of magnitude higher than that of the guided topological mode with $w=+1$.

		\begin{figure*}
		\centering
		\includegraphics[width=1\textwidth]{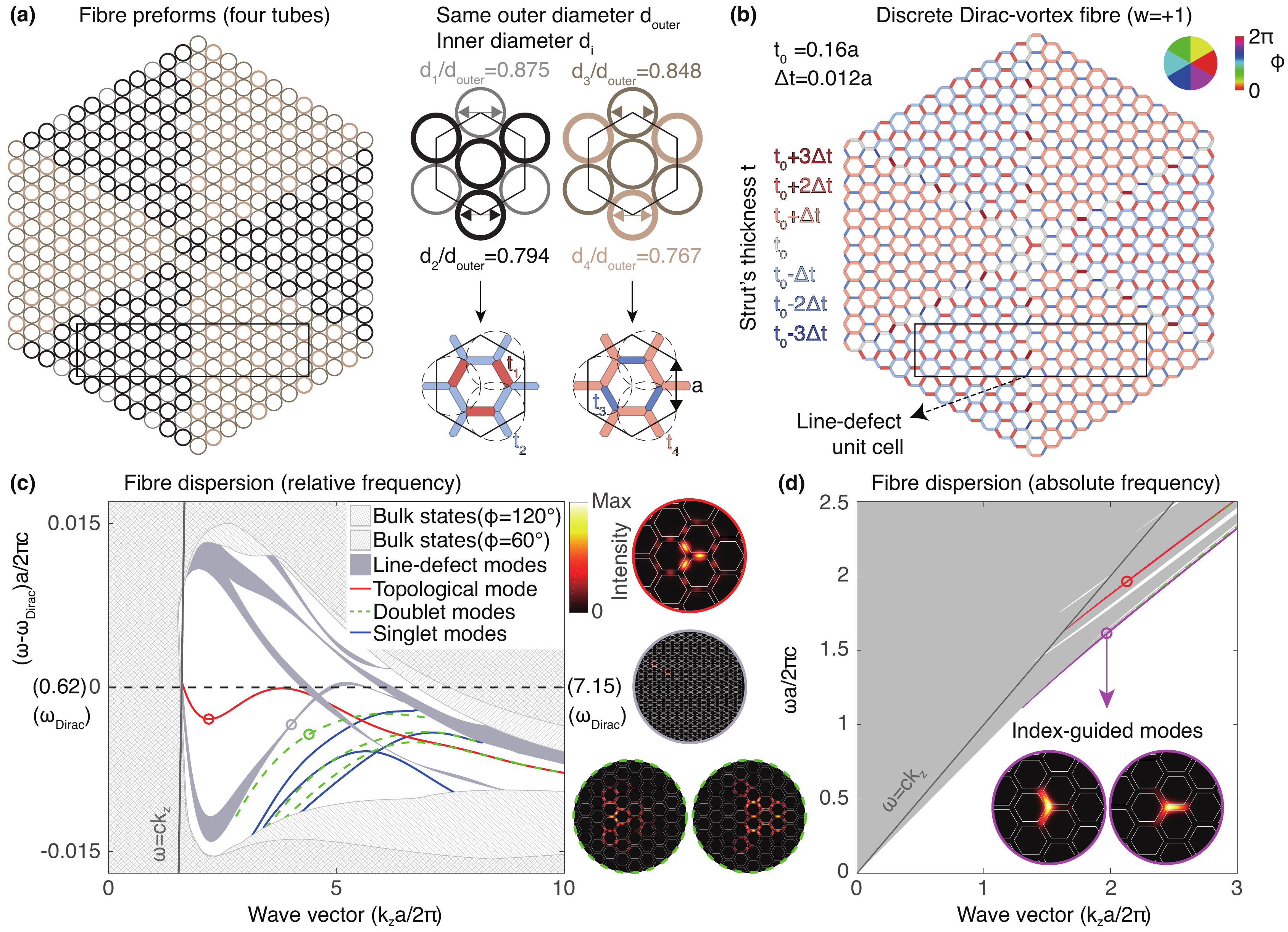}
		\caption{Dirac-vortex fibre obtained by discrete Kekul$\rm\acute{e}$ modulations. (a) Fibre preform with four capillary silica tubes of the same outer diameter. (b) Fibre structure, having seven different strut widths, resulting from the preform in (a). The black rectangle is the unit cell for the line-defect modes. (c) Band diagram with the frequency referenced to the original nodal line of the unmodulated lattice. The higher-order modes are plotted with blue and green lines according their degeneracy. Insets show the mode profiles~($\hat{\boldsymbol{ z}}\cdot \rm Re[\boldsymbol{ E}^{*} \times \boldsymbol{ H}]$). (d) Purple dispersion and mode profiles of the polarization-degenerate index-guided fundamental mode. Other index-guide modes are plotted with green dashes.
}
		\label{fig::discrete} 
	\end{figure*}

\subsection{Arbitrary degenerate modes}

One main topological feature of the Dirac-vortex fibre is the ease of creating multiple near-degenerate modes by simply increasing the winding number~($w$) of the vortex~\cite{gao2020dirac}. 
We demonstrate this in Fig.~\ref{fig::multimode} for $w=+1,+2,+3$, where $w$ is the topological invariant of the system and can be an arbitrary integer. The sign of $w$ determines where the field peaks around one of the two sublattices of a honeycomb lattice. The two sublattices can be viewed as the two joints, each having three struts, in the primitive cell of Fig.~\ref{fig::nodalline}(a). An example of $w=-1$ is presented in Supplementary Part I.

In these multimode examples, we decrease the modulation amplitude~(as well as the bandgap size) to eliminate the local-defect modes in the momentum range plotted. Additionally, we keep the core centre at the $C_{3v}$ centre of the $w=+1$ vortex. If we choose a w-dependent centre, then any vortex can be $C_{3v}$ symmetric, so some pairs of fibre dispersions will be rigorously degenerate due to the doublet representation of the point group.

In principle, a continuous, single-mode or multimode, Dirac-vortex PCF can be fabricated either from 3D-printed preforms or by the stack-and-draw method with over a hundred tubes with different tube thicknesses. Neither of the two solutions are convenient. Therefore, we present the discrete version of the fibre design.

\subsection{Discrete modulation with four tubes}
	
We discretize the continuous vortex design~[Fig.\ref{fig::continuous}(e)] into six discrete modulation phase angles~($\phi$).
Due to the $C_{3v}$ symmetry of the cross-section, only two phases of the six have distinct structures. The three phases that differ by 120$^\circ$ are actually identical in the unit-cell structure (but with different origins and orientations). As a result, we only need four tubes to stack-and-draw the Dirac-vortex PCF, which is very reasonable for fabrication. Another discrete version is presented in Supplementary Part II using four strut thicknesses.

As shown in Fig.~\ref{fig::discrete}(a), the four silica capillary tubes have the same outer diameter $d_{\rm{outer}}$ to maintain the lattice but different inner diameters $d_i$ for modulation. The four exact ratios of $d_i/d_{\rm{outer}}$ can be determined from Eq.~\ref{eq::ratio}, satisfying the thickness and area proportionality.
		\begin{equation}  
		\begin{cases}
		\frac{d_{\rm{outer}}-(d_{1,4}+d_{2,3})/2}{t_{2,4}}&=\frac{d_{\rm{outer}}-d_{2,3}}{t_{1,3}}\\
\frac{\pi(3d_{\rm{outer}}^{2}-d_{1,4}^{2}-2d_{2,3}^{2})}{4d_{\rm{outer}}^{2}}
&= \frac{3\sqrt{3}t_{0}(a-t_{0}/2)}{a^{2}}
		\end{cases}
		\label{eq::ratio}
		\end{equation}

The resulting discrete Dirac-vortex fibre and its band structure are plotted in Fig.~\ref{fig::discrete}(b) and (c). Compared to the continuous version in Fig.~\ref{fig::continuous}(e) and (f), the structural nonuniformity now only exists at the six identical interfaces~(also determined by Eq.~\ref{eq::ratio}) between the two distinct lattices~\cite{wu2015scheme}, at which the extra defect modes can locate. In addition to these line-defect modes inside the bandgap, there are also higher-order vortex modes plotted in green and blue lines, according to their $C_{3v}$ point group representations. The mode profiles of the higher-order modes are much larger than that of the topological mode.

We have been focusing on the modes inside the bandgap. However, as shown in Fig.~\ref{fig::discrete}(d)~[also in Fig.~\ref{fig::continuous}(e, f)], there could also be index-guided modes in the Dirac-vortex fibre whose frequency is the lowest for a common wave vector. Index-guided modes exist wherever there is a sharp local maximum of the strut thickness, equivalent to a local rise of the effective refractive index. The loss of the index-guided modes is usually much lower than that of the in-gap modes.

\subsection{Nonzero vortex size with continuous modulation}
	
To operate at the topological mid-gap dispersion, we have to remove the index-guided modes in the design. We do this by enlarging the vortex size and smoothing the sharp lattice~(strut) interfaces at the fibre core. The smooth modulation envelope function is $\Delta t(r)=\rm{tanh}(r/R)$, where $r$ is the radial distance and $R$ is the vortex radius of choice, similar to the 2D cavity design in Ref.~\cite{gao2020dirac}. Note that in the previous examples of this paper, the vortex size is $R=0$.

\begin{figure*}
	\centering
	\includegraphics[width=\textwidth]{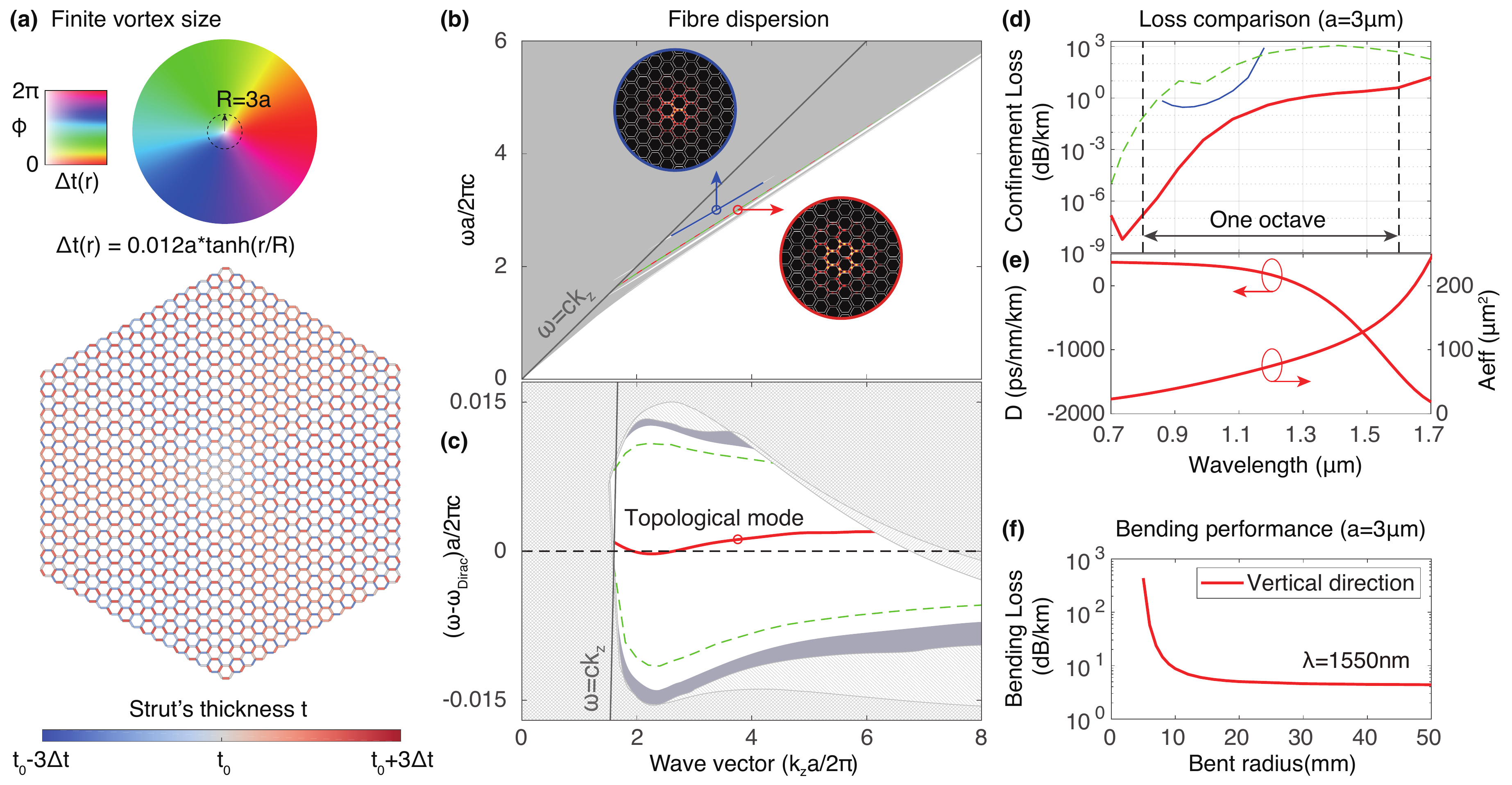}
	\caption{Octave SPSM in a continuous Dirac-vortex fibre with a finite vortex size. (a) Fibre structure with sixteen cladding periods in radius. The colour wheel represents the phase and amplitude of the generalized $\rm{Kekul\acute{e}}$ modulation. (b) Complete fibre dispersion in absolute frequency. First topological mode~(red line) and higher-order doublet modes~(green dotted line) in the first topological bandgap, as well as a second topological mode~(blue line) in the second topological bandgap at higher frequency. The mode profiles~($\hat{\boldsymbol{ z}}\cdot \rm Re[\boldsymbol{ E}^{*} \times \boldsymbol{ H}]$) of the two topological modes are shown in the insets, circled with different colours for clarity. (c) Fibre dispersion in frequency relative to the original nodal-line frequency. (d) Confinement losses of the guided modes. (e) Dispersion parameter and effective area of the first topological mode. (f) Bending loss of the first topological mode at $\lambda=1550$ $nm$.
	}
	\label{fig::SPSM}
	\end{figure*}

	The fibre cross-section with vortex size $R=3a$ is shown in Fig.~\ref{fig::SPSM}(a), and the corresponding band structure is shown in Fig.~\ref{fig::SPSM}(b) and (c). There are no index-guided modes at the bottom of the bands. For completeness, we also plot the blue mid-gap dispersion inside the second topological bandgap, originating from a higher frequency Dirac point (nodal line) illustrated in Supplementary Part III. We did not discuss it in the previous examples because it has a much higher loss than the red mid-gap modes in the first topological gap, as shown in Fig.~\ref{fig::SPSM}(d).

\subsection{Octave SPSM}
The continuously modulated Dirac-vortex PCF with a finite vortex diameter~($2R=6a$) has an SPSM. We evaluate its potential performance in terms of the confinement loss, dispersion parameter, effective area and bending loss in Fig.~\ref{fig::SPSM}(d,e,f).
We choose the wave vector~($k_{z}a/2\pi$) range from 1.8 to 5.7 and the period $a=3$ $\mu m$, so that the corresponding wavelength ranges from $1700$ $nm$ to $700$ $nm$ --- the low-loss window of silica.

The modes with the lowest confinement losses are plotted in Fig.~\ref{fig::SPSM}(d), computed with absorbing boundary conditions. The loss of the topological mode~(red) is always the lowest for the whole wavelength range over one octave.
Most of its confinement loss is less than 1 $dB/km$ with sixteen layers of air holes along the fibre radius~[Fig.~\ref{fig::SPSM}(a)].

The dispersion parameter $D$ of the topological mode can take large negative values of -1500 $ps/nm/km$, and the zero-dispersion wavelength is approximately 1.3 $\mu m$. The effective area~($A_{\rm{eff}}$ defined in Ref.~\cite{joannopoulos2008molding}) is 23 $\mu m^{2}$ at a 0.7 $\mu m$ wavelength and increases to 243 $\mu m^{2}$ at a 1.7 $\mu m$ wavelength. The bending loss is calculated using a conformal transformation analysis~\cite{heiblum1975analysis} with an equivalent refractive index profile  $n_{eq}(x,y) = n(x,y){\rm exp}(x/X)$~\cite{beravat2016twist} for the bending radius $X$.
The results in Fig.~\ref{fig::SPSM}(f) show that the fibre can be bent to a radius as small as 15 $mm$ without experiencing significant losses. These specifications of the Dirac-vortex PCF are similar to those of previous PCFs~\cite{finazzi2003small,tsuchida2005design}, with the key difference being that the topological mode is a single-polarization mode.

	\section{Discussion}
	
	We numerically investigate the Dirac-vortex topological PCF in terms of its principle, construction, and potential performance. This fibre could be made with the standard stack-and-draw process with silica glass tubes or 3D-printed preforms. The design is tolerant of structural details such as the interface curvatures, not considered in the main text but discussed in Supplementary Part IV. It is also possible to move the topological dispersion above the light line, as discussed in Supplementary Part V.
	
Similar to the case of a three-dimensional one-way fibre~\cite{lu2018topological}, where the topological invariant is a four-dimensional second Chern number, the topological invariant of the Dirac-vortex PCF can be written as a three-dimensional integral~(equation 4.2 in Ref.~\cite{teo2010topological} under chiral symmetry) where the coordinates are $k_x, k_y$ and the polar angle (a form of synthetic dimension ~\cite{yuan2018synthetic}).

The advantage of the Dirac-vortex PCF over previous fibres is the ability to guide any number of near-degenerate modes at will. The single-mode design provides an SPSM fibre with an octave bandwidth. The effective mode area is easily tuneable by changing the vortex size~($R$). This work suggests PCFs as a new platform for topological photonics.
	
	\section{Acknowledgements}
	
We thank Wei Ding, YingYing Wang, Xiaomei Gao, Zhong Wang and Guoqing Chang for helpful discussions. This work was supported by the National Key R\&D Program of China~(2017YFA0303800, 2016YFA0302400), the Natural Science Foundation of China (11721404), the Strategic Priority Research Program~(XDB33000000) and the international partnership program~(112111KYSB20200024) of the Chinese Academy of Sciences, and the Beijing Natural Science Foundation~(Z200008).

\section{Author contributions}
Both authors contributed to the design of the fibre and the writing of the manuscript. H.L. performed the numerical simulations.

\section{Conflict of interest}
The authors declare that they have no conflict of interest. 

\newpage
	\bibliography{citationlist}

\begin{thebibliography}{40}%
\makeatletter
\providecommand \@ifxundefined [1]{%
 \@ifx{#1\undefined}
}%
\providecommand \@ifnum [1]{%
 \ifnum #1\expandafter \@firstoftwo
 \else \expandafter \@secondoftwo
 \fi
}%
\providecommand \@ifx [1]{%
 \ifx #1\expandafter \@firstoftwo
 \else \expandafter \@secondoftwo
 \fi
}%
\providecommand \natexlab [1]{#1}%
\providecommand \enquote  [1]{``#1''}%
\providecommand \bibnamefont  [1]{#1}%
\providecommand \bibfnamefont [1]{#1}%
\providecommand \citenamefont [1]{#1}%
\providecommand \href@noop [0]{\@secondoftwo}%
\providecommand \href [0]{\begingroup \@sanitize@url \@href}%
\providecommand \@href[1]{\@@startlink{#1}\@@href}%
\providecommand \@@href[1]{\endgroup#1\@@endlink}%
\providecommand \@sanitize@url [0]{\catcode `\\12\catcode `\$12\catcode
  `\&12\catcode `\#12\catcode `\^12\catcode `\_12\catcode `\%12\relax}%
\providecommand \@@startlink[1]{}%
\providecommand \@@endlink[0]{}%
\providecommand \url  [0]{\begingroup\@sanitize@url \@url }%
\providecommand \@url [1]{\endgroup\@href {#1}{\urlprefix }}%
\providecommand \urlprefix  [0]{URL }%
\providecommand \Eprint [0]{\href }%
\providecommand \doibase [0]{http://dx.doi.org/}%
\providecommand \selectlanguage [0]{\@gobble}%
\providecommand \bibinfo  [0]{\@secondoftwo}%
\providecommand \bibfield  [0]{\@secondoftwo}%
\providecommand \translation [1]{[#1]}%
\providecommand \BibitemOpen [0]{}%
\providecommand \bibitemStop [0]{}%
\providecommand \bibitemNoStop [0]{.\EOS\space}%
\providecommand \EOS [0]{\spacefactor3000\relax}%
\providecommand \BibitemShut  [1]{\csname bibitem#1\endcsname}%
\let\auto@bib@innerbib\@empty
\bibitem [{\citenamefont {Lu}\ \emph {et~al.}(2014)\citenamefont {Lu},
  \citenamefont {Joannopoulos},\ and\ \citenamefont
  {Solja{\v{c}}i{\'c}}}]{lu2014topological}%
  \BibitemOpen
  \bibfield  {author} {\bibinfo {author} {\bibfnamefont {Ling}\ \bibnamefont
  {Lu}}, \bibinfo {author} {\bibfnamefont {John~D}\ \bibnamefont
  {Joannopoulos}}, \ and\ \bibinfo {author} {\bibfnamefont {Marin}\
  \bibnamefont {Solja{\v{c}}i{\'c}}},\ }\bibfield  {title} {\enquote {\bibinfo
  {title} {Topological photonics},}\ }\href@noop {} {\bibfield  {journal}
  {\bibinfo  {journal} {Nature Photonics}\ }\textbf {\bibinfo {volume} {8}},\
  \bibinfo {pages} {821--829} (\bibinfo {year} {2014})}\BibitemShut {NoStop}%
\bibitem [{\citenamefont {Khanikaev}\ and\ \citenamefont
  {Shvets}(2017)}]{khanikaev2017two}%
  \BibitemOpen
  \bibfield  {author} {\bibinfo {author} {\bibfnamefont {Alexander~B}\
  \bibnamefont {Khanikaev}}\ and\ \bibinfo {author} {\bibfnamefont {Gennady}\
  \bibnamefont {Shvets}},\ }\bibfield  {title} {\enquote {\bibinfo {title}
  {Two-dimensional topological photonics},}\ }\href@noop {} {\bibfield
  {journal} {\bibinfo  {journal} {Nature Photonics}\ }\textbf {\bibinfo
  {volume} {11}},\ \bibinfo {pages} {763} (\bibinfo {year} {2017})}\BibitemShut
  {NoStop}%
\bibitem [{\citenamefont {Ozawa}\ \emph {et~al.}(2019)\citenamefont {Ozawa},
  \citenamefont {Price}, \citenamefont {Amo}, \citenamefont {Goldman},
  \citenamefont {Hafezi}, \citenamefont {Lu}, \citenamefont {Rechtsman},
  \citenamefont {Schuster}, \citenamefont {Simon}, \citenamefont {Zilberberg}
  \emph {et~al.}}]{ozawa2019topological}%
  \BibitemOpen
  \bibfield  {author} {\bibinfo {author} {\bibfnamefont {Tomoki}\ \bibnamefont
  {Ozawa}}, \bibinfo {author} {\bibfnamefont {Hannah~M}\ \bibnamefont {Price}},
  \bibinfo {author} {\bibfnamefont {Alberto}\ \bibnamefont {Amo}}, \bibinfo
  {author} {\bibfnamefont {Nathan}\ \bibnamefont {Goldman}}, \bibinfo {author}
  {\bibfnamefont {Mohammad}\ \bibnamefont {Hafezi}}, \bibinfo {author}
  {\bibfnamefont {Ling}\ \bibnamefont {Lu}}, \bibinfo {author} {\bibfnamefont
  {Mikael~C}\ \bibnamefont {Rechtsman}}, \bibinfo {author} {\bibfnamefont
  {David}\ \bibnamefont {Schuster}}, \bibinfo {author} {\bibfnamefont
  {Jonathan}\ \bibnamefont {Simon}}, \bibinfo {author} {\bibfnamefont {Oded}\
  \bibnamefont {Zilberberg}},  \emph {et~al.},\ }\bibfield  {title} {\enquote
  {\bibinfo {title} {Topological photonics},}\ }\href@noop {} {\bibfield
  {journal} {\bibinfo  {journal} {Reviews of Modern Physics}\ }\textbf
  {\bibinfo {volume} {91}},\ \bibinfo {pages} {015006} (\bibinfo {year}
  {2019})}\BibitemShut {NoStop}%
\bibitem [{\citenamefont {Lu}\ \emph {et~al.}(2018)\citenamefont {Lu},
  \citenamefont {Gao},\ and\ \citenamefont {Wang}}]{lu2018topological}%
  \BibitemOpen
  \bibfield  {author} {\bibinfo {author} {\bibfnamefont {Ling}\ \bibnamefont
  {Lu}}, \bibinfo {author} {\bibfnamefont {Haozhe}\ \bibnamefont {Gao}}, \ and\
  \bibinfo {author} {\bibfnamefont {Zhong}\ \bibnamefont {Wang}},\ }\bibfield
  {title} {\enquote {\bibinfo {title} {Topological one-way fiber of second
  chern number},}\ }\href@noop {} {\bibfield  {journal} {\bibinfo  {journal}
  {Nature communications}\ }\textbf {\bibinfo {volume} {9}},\ \bibinfo {pages}
  {1--7} (\bibinfo {year} {2018})}\BibitemShut {NoStop}%
\bibitem [{\citenamefont {Pilozzi}\ \emph {et~al.}(2020)\citenamefont
  {Pilozzi}, \citenamefont {Leykam}, \citenamefont {Chen},\ and\ \citenamefont
  {Conti}}]{pilozzi2020topological}%
  \BibitemOpen
  \bibfield  {author} {\bibinfo {author} {\bibfnamefont {L}~\bibnamefont
  {Pilozzi}}, \bibinfo {author} {\bibfnamefont {Daniel}\ \bibnamefont
  {Leykam}}, \bibinfo {author} {\bibfnamefont {Zhigang}\ \bibnamefont {Chen}},
  \ and\ \bibinfo {author} {\bibfnamefont {Claudio}\ \bibnamefont {Conti}},\
  }\bibfield  {title} {\enquote {\bibinfo {title} {Topological photonic crystal
  fibers and ring resonators.}}\ }\href@noop {} {\bibfield  {journal} {\bibinfo
   {journal} {Optics Letters}\ }\textbf {\bibinfo {volume} {45}},\ \bibinfo
  {pages} {1415--1418} (\bibinfo {year} {2020})}\BibitemShut {NoStop}%
\bibitem [{\citenamefont {Gao}\ \emph {et~al.}(2020)\citenamefont {Gao},
  \citenamefont {Yang}, \citenamefont {Lin}, \citenamefont {Zhang},
  \citenamefont {Li}, \citenamefont {Bo}, \citenamefont {Wang},\ and\
  \citenamefont {Lu}}]{gao2020dirac}%
  \BibitemOpen
  \bibfield  {author} {\bibinfo {author} {\bibfnamefont {Xiaomei}\ \bibnamefont
  {Gao}}, \bibinfo {author} {\bibfnamefont {Lechen}\ \bibnamefont {Yang}},
  \bibinfo {author} {\bibfnamefont {Hao}\ \bibnamefont {Lin}}, \bibinfo
  {author} {\bibfnamefont {Lang}\ \bibnamefont {Zhang}}, \bibinfo {author}
  {\bibfnamefont {Jiafang}\ \bibnamefont {Li}}, \bibinfo {author}
  {\bibfnamefont {Fang}\ \bibnamefont {Bo}}, \bibinfo {author} {\bibfnamefont
  {Zhong}\ \bibnamefont {Wang}}, \ and\ \bibinfo {author} {\bibfnamefont
  {Ling}\ \bibnamefont {Lu}},\ }\bibfield  {title} {\enquote {\bibinfo {title}
  {Dirac-vortex topological cavities},}\ }\href@noop {} {\bibfield  {journal}
  {\bibinfo  {journal} {Nature Nanotechnology}\ ,\ \bibinfo {pages} {1--7}}
  (\bibinfo {year} {2020})}\BibitemShut {NoStop}%
\bibitem [{\citenamefont {Jackiw}\ and\ \citenamefont
  {Rossi}(1981)}]{jackiw1981zero}%
  \BibitemOpen
  \bibfield  {author} {\bibinfo {author} {\bibfnamefont {R}~\bibnamefont
  {Jackiw}}\ and\ \bibinfo {author} {\bibfnamefont {Paolo}\ \bibnamefont
  {Rossi}},\ }\bibfield  {title} {\enquote {\bibinfo {title} {Zero modes of the
  vortex-fermion system},}\ }\href@noop {} {\bibfield  {journal} {\bibinfo
  {journal} {Nuclear Physics B}\ }\textbf {\bibinfo {volume} {190}},\ \bibinfo
  {pages} {681--691} (\bibinfo {year} {1981})}\BibitemShut {NoStop}%
\bibitem [{\citenamefont {Hou}\ \emph {et~al.}(2007)\citenamefont {Hou},
  \citenamefont {Chamon},\ and\ \citenamefont {Mudry}}]{hou2007electron}%
  \BibitemOpen
  \bibfield  {author} {\bibinfo {author} {\bibfnamefont {Chang-Yu}\
  \bibnamefont {Hou}}, \bibinfo {author} {\bibfnamefont {Claudio}\ \bibnamefont
  {Chamon}}, \ and\ \bibinfo {author} {\bibfnamefont {Christopher}\
  \bibnamefont {Mudry}},\ }\bibfield  {title} {\enquote {\bibinfo {title}
  {Electron fractionalization in two-dimensional graphenelike structures},}\
  }\href@noop {} {\bibfield  {journal} {\bibinfo  {journal} {Physical review
  letters}\ }\textbf {\bibinfo {volume} {98}},\ \bibinfo {pages} {186809}
  (\bibinfo {year} {2007})}\BibitemShut {NoStop}%
\bibitem [{\citenamefont {Iadecola}\ \emph {et~al.}(2016)\citenamefont
  {Iadecola}, \citenamefont {Schuster},\ and\ \citenamefont
  {Chamon}}]{iadecola2016non}%
  \BibitemOpen
  \bibfield  {author} {\bibinfo {author} {\bibfnamefont {Thomas}\ \bibnamefont
  {Iadecola}}, \bibinfo {author} {\bibfnamefont {Thomas}\ \bibnamefont
  {Schuster}}, \ and\ \bibinfo {author} {\bibfnamefont {Claudio}\ \bibnamefont
  {Chamon}},\ }\bibfield  {title} {\enquote {\bibinfo {title} {Non-abelian
  braiding of light},}\ }\href@noop {} {\bibfield  {journal} {\bibinfo
  {journal} {Physical Review Letters}\ }\textbf {\bibinfo {volume} {117}},\
  \bibinfo {pages} {073901} (\bibinfo {year} {2016})}\BibitemShut {NoStop}%
\bibitem [{\citenamefont {Menssen}\ \emph {et~al.}(2020)\citenamefont
  {Menssen}, \citenamefont {Guan}, \citenamefont {Felce}, \citenamefont
  {Booth},\ and\ \citenamefont {Walmsley}}]{menssen2020photonic}%
  \BibitemOpen
  \bibfield  {author} {\bibinfo {author} {\bibfnamefont {Adrian~J}\
  \bibnamefont {Menssen}}, \bibinfo {author} {\bibfnamefont {Jun}\ \bibnamefont
  {Guan}}, \bibinfo {author} {\bibfnamefont {David}\ \bibnamefont {Felce}},
  \bibinfo {author} {\bibfnamefont {Martin~J}\ \bibnamefont {Booth}}, \ and\
  \bibinfo {author} {\bibfnamefont {Ian~A}\ \bibnamefont {Walmsley}},\
  }\bibfield  {title} {\enquote {\bibinfo {title} {Photonic topological mode
  bound to a vortex},}\ }\href@noop {} {\bibfield  {journal} {\bibinfo
  {journal} {Physical Review Letters}\ }\textbf {\bibinfo {volume} {125}},\
  \bibinfo {pages} {117401} (\bibinfo {year} {2020})}\BibitemShut {NoStop}%
\bibitem [{\citenamefont {Noh}\ \emph {et~al.}(2020)\citenamefont {Noh},
  \citenamefont {Schuster}, \citenamefont {Iadecola}, \citenamefont {Huang},
  \citenamefont {Wang}, \citenamefont {Chen}, \citenamefont {Chamon},\ and\
  \citenamefont {Rechtsman}}]{noh2020braiding}%
  \BibitemOpen
  \bibfield  {author} {\bibinfo {author} {\bibfnamefont {Jiho}\ \bibnamefont
  {Noh}}, \bibinfo {author} {\bibfnamefont {Thomas}\ \bibnamefont {Schuster}},
  \bibinfo {author} {\bibfnamefont {Thomas}\ \bibnamefont {Iadecola}}, \bibinfo
  {author} {\bibfnamefont {Sheng}\ \bibnamefont {Huang}}, \bibinfo {author}
  {\bibfnamefont {Mohan}\ \bibnamefont {Wang}}, \bibinfo {author}
  {\bibfnamefont {Kevin~P}\ \bibnamefont {Chen}}, \bibinfo {author}
  {\bibfnamefont {Claudio}\ \bibnamefont {Chamon}}, \ and\ \bibinfo {author}
  {\bibfnamefont {Mikael~C}\ \bibnamefont {Rechtsman}},\ }\bibfield  {title}
  {\enquote {\bibinfo {title} {Braiding photonic topological zero modes},}\
  }\href@noop {} {\bibfield  {journal} {\bibinfo  {journal} {Nature Physics}\
  }\textbf {\bibinfo {volume} {16}},\ \bibinfo {pages} {989--993} (\bibinfo
  {year} {2020})}\BibitemShut {NoStop}%
\bibitem [{\citenamefont {Gao}\ \emph {et~al.}(2019)\citenamefont {Gao},
  \citenamefont {Torrent}, \citenamefont {Cervera}, \citenamefont {San-Jose},
  \citenamefont {S{\'a}nchez-Dehesa},\ and\ \citenamefont
  {Christensen}}]{gao2019majorana}%
  \BibitemOpen
  \bibfield  {author} {\bibinfo {author} {\bibfnamefont {Penglin}\ \bibnamefont
  {Gao}}, \bibinfo {author} {\bibfnamefont {Daniel}\ \bibnamefont {Torrent}},
  \bibinfo {author} {\bibfnamefont {Francisco}\ \bibnamefont {Cervera}},
  \bibinfo {author} {\bibfnamefont {Pablo}\ \bibnamefont {San-Jose}}, \bibinfo
  {author} {\bibfnamefont {Jos{\'e}}\ \bibnamefont {S{\'a}nchez-Dehesa}}, \
  and\ \bibinfo {author} {\bibfnamefont {Johan}\ \bibnamefont {Christensen}},\
  }\bibfield  {title} {\enquote {\bibinfo {title} {Majorana-like zero modes in
  kekul{\'e} distorted sonic lattices},}\ }\href@noop {} {\bibfield  {journal}
  {\bibinfo  {journal} {Physical review letters}\ }\textbf {\bibinfo {volume}
  {123}},\ \bibinfo {pages} {196601} (\bibinfo {year} {2019})}\BibitemShut
  {NoStop}%
\bibitem [{\citenamefont {Chen}\ \emph {et~al.}(2019)\citenamefont {Chen},
  \citenamefont {Lera}, \citenamefont {Chaunsali}, \citenamefont {Torrent},
  \citenamefont {Alvarez}, \citenamefont {Yang}, \citenamefont {San-Jose},\
  and\ \citenamefont {Christensen}}]{chen2019mechanical}%
  \BibitemOpen
  \bibfield  {author} {\bibinfo {author} {\bibfnamefont {Chun-Wei}\
  \bibnamefont {Chen}}, \bibinfo {author} {\bibfnamefont {Natalia}\
  \bibnamefont {Lera}}, \bibinfo {author} {\bibfnamefont {Rajesh}\ \bibnamefont
  {Chaunsali}}, \bibinfo {author} {\bibfnamefont {Daniel}\ \bibnamefont
  {Torrent}}, \bibinfo {author} {\bibfnamefont {Jose~Vicente}\ \bibnamefont
  {Alvarez}}, \bibinfo {author} {\bibfnamefont {Jinkyu}\ \bibnamefont {Yang}},
  \bibinfo {author} {\bibfnamefont {Pablo}\ \bibnamefont {San-Jose}}, \ and\
  \bibinfo {author} {\bibfnamefont {Johan}\ \bibnamefont {Christensen}},\
  }\bibfield  {title} {\enquote {\bibinfo {title} {Mechanical analogue of a
  majorana bound state},}\ }\href@noop {} {\bibfield  {journal} {\bibinfo
  {journal} {Advanced Materials}\ }\textbf {\bibinfo {volume} {31}},\ \bibinfo
  {pages} {1904386} (\bibinfo {year} {2019})}\BibitemShut {NoStop}%
\bibitem [{\citenamefont {Okoshi}\ and\ \citenamefont
  {Oyamada}(1980)}]{okoshi1980single}%
  \BibitemOpen
  \bibfield  {author} {\bibinfo {author} {\bibfnamefont {T}~\bibnamefont
  {Okoshi}}\ and\ \bibinfo {author} {\bibfnamefont {K}~\bibnamefont
  {Oyamada}},\ }\bibfield  {title} {\enquote {\bibinfo {title}
  {Single-polarisation single-mode optical fibre with refractive-index pits on
  both sides of core},}\ }\href@noop {} {\bibfield  {journal} {\bibinfo
  {journal} {Electronics Letters}\ }\textbf {\bibinfo {volume} {16}},\ \bibinfo
  {pages} {712--713} (\bibinfo {year} {1980})}\BibitemShut {NoStop}%
\bibitem [{\citenamefont {Eickhoff}(1982)}]{eickhoff1982stress}%
  \BibitemOpen
  \bibfield  {author} {\bibinfo {author} {\bibfnamefont {W}~\bibnamefont
  {Eickhoff}},\ }\bibfield  {title} {\enquote {\bibinfo {title} {Stress-induced
  single-polarization single-mode fiber},}\ }\href@noop {} {\bibfield
  {journal} {\bibinfo  {journal} {Optics letters}\ }\textbf {\bibinfo {volume}
  {7}},\ \bibinfo {pages} {629--631} (\bibinfo {year} {1982})}\BibitemShut
  {NoStop}%
\bibitem [{\citenamefont {Simpson}\ \emph {et~al.}(1983)\citenamefont
  {Simpson}, \citenamefont {Stolen}, \citenamefont {Sears}, \citenamefont
  {Pleibel}, \citenamefont {MacChesney},\ and\ \citenamefont
  {Howard}}]{simpson1983single}%
  \BibitemOpen
  \bibfield  {author} {\bibinfo {author} {\bibfnamefont {J}~\bibnamefont
  {Simpson}}, \bibinfo {author} {\bibfnamefont {R}~\bibnamefont {Stolen}},
  \bibinfo {author} {\bibfnamefont {F}~\bibnamefont {Sears}}, \bibinfo {author}
  {\bibfnamefont {William}\ \bibnamefont {Pleibel}}, \bibinfo {author}
  {\bibfnamefont {J}~\bibnamefont {MacChesney}}, \ and\ \bibinfo {author}
  {\bibfnamefont {Richard}\ \bibnamefont {Howard}},\ }\bibfield  {title}
  {\enquote {\bibinfo {title} {A single-polarization fiber},}\ }\href@noop {}
  {\bibfield  {journal} {\bibinfo  {journal} {Journal of Lightwave Technology}\
  }\textbf {\bibinfo {volume} {1}},\ \bibinfo {pages} {370--374} (\bibinfo
  {year} {1983})}\BibitemShut {NoStop}%
\bibitem [{\citenamefont {Kubota}\ \emph {et~al.}(2004)\citenamefont {Kubota},
  \citenamefont {Kawanishi}, \citenamefont {Koyanagi}, \citenamefont {Tanaka},\
  and\ \citenamefont {Yamaguchi}}]{kubota2004absolutely}%
  \BibitemOpen
  \bibfield  {author} {\bibinfo {author} {\bibfnamefont {Hiirokazu}\
  \bibnamefont {Kubota}}, \bibinfo {author} {\bibfnamefont {Satoki}\
  \bibnamefont {Kawanishi}}, \bibinfo {author} {\bibfnamefont {Shigeki}\
  \bibnamefont {Koyanagi}}, \bibinfo {author} {\bibfnamefont {Masatoshi}\
  \bibnamefont {Tanaka}}, \ and\ \bibinfo {author} {\bibfnamefont
  {Shyunichiro}\ \bibnamefont {Yamaguchi}},\ }\bibfield  {title} {\enquote
  {\bibinfo {title} {Absolutely single polarization photonic crystal fiber},}\
  }\href@noop {} {\bibfield  {journal} {\bibinfo  {journal} {IEEE Photonics
  Technology Letters}\ }\textbf {\bibinfo {volume} {16}},\ \bibinfo {pages}
  {182--184} (\bibinfo {year} {2004})}\BibitemShut {NoStop}%
\bibitem [{\citenamefont {Folkenberg}\ \emph {et~al.}(2005)\citenamefont
  {Folkenberg}, \citenamefont {Nielsen},\ and\ \citenamefont
  {Jakobsen}}]{folkenberg2005broadband}%
  \BibitemOpen
  \bibfield  {author} {\bibinfo {author} {\bibfnamefont {JR}~\bibnamefont
  {Folkenberg}}, \bibinfo {author} {\bibfnamefont {MD}~\bibnamefont {Nielsen}},
  \ and\ \bibinfo {author} {\bibfnamefont {C}~\bibnamefont {Jakobsen}},\
  }\bibfield  {title} {\enquote {\bibinfo {title} {Broadband
  single-polarization photonic crystal fiber},}\ }\href@noop {} {\bibfield
  {journal} {\bibinfo  {journal} {Optics letters}\ }\textbf {\bibinfo {volume}
  {30}},\ \bibinfo {pages} {1446--1448} (\bibinfo {year} {2005})}\BibitemShut
  {NoStop}%
\bibitem [{\citenamefont {Lee}\ \emph {et~al.}(2008)\citenamefont {Lee},
  \citenamefont {Avniel},\ and\ \citenamefont {Johnson}}]{lee2008design}%
  \BibitemOpen
  \bibfield  {author} {\bibinfo {author} {\bibfnamefont {Karen~KY}\
  \bibnamefont {Lee}}, \bibinfo {author} {\bibfnamefont {Yehuda}\ \bibnamefont
  {Avniel}}, \ and\ \bibinfo {author} {\bibfnamefont {Steven~G}\ \bibnamefont
  {Johnson}},\ }\bibfield  {title} {\enquote {\bibinfo {title} {Design
  strategies and rigorous conditions for single-polarization single-mode
  waveguides},}\ }\href@noop {} {\bibfield  {journal} {\bibinfo  {journal}
  {Optics express}\ }\textbf {\bibinfo {volume} {16}},\ \bibinfo {pages}
  {15170--15184} (\bibinfo {year} {2008})}\BibitemShut {NoStop}%
\bibitem [{\citenamefont {Ferrando}\ and\ \citenamefont
  {Miret}(2001)}]{ferrando2001single}%
  \BibitemOpen
  \bibfield  {author} {\bibinfo {author} {\bibfnamefont {Albert}\ \bibnamefont
  {Ferrando}}\ and\ \bibinfo {author} {\bibfnamefont {Juan~Jos{\'e}}\
  \bibnamefont {Miret}},\ }\bibfield  {title} {\enquote {\bibinfo {title}
  {Single-polarization single-mode intraband guidance in supersquare photonic
  crystals fibers},}\ }\href@noop {} {\bibfield  {journal} {\bibinfo  {journal}
  {Applied Physics Letters}\ }\textbf {\bibinfo {volume} {78}},\ \bibinfo
  {pages} {3184--3186} (\bibinfo {year} {2001})}\BibitemShut {NoStop}%
\bibitem [{\citenamefont {Eguchi}\ and\ \citenamefont
  {Tsuji}(2012)}]{eguchi2012single}%
  \BibitemOpen
  \bibfield  {author} {\bibinfo {author} {\bibfnamefont {Masashi}\ \bibnamefont
  {Eguchi}}\ and\ \bibinfo {author} {\bibfnamefont {Yasuhide}\ \bibnamefont
  {Tsuji}},\ }\bibfield  {title} {\enquote {\bibinfo {title}
  {Single-polarization elliptical-hole lattice core photonic-bandgap fiber},}\
  }\href@noop {} {\bibfield  {journal} {\bibinfo  {journal} {Journal of
  lightwave technology}\ }\textbf {\bibinfo {volume} {31}},\ \bibinfo {pages}
  {177--182} (\bibinfo {year} {2012})}\BibitemShut {NoStop}%
\bibitem [{\citenamefont {Szpulak}\ \emph {et~al.}(2007)\citenamefont
  {Szpulak}, \citenamefont {Martynkien}, \citenamefont {Olszewski},
  \citenamefont {Urbanczyk}, \citenamefont {Nasilowski}, \citenamefont
  {Berghmans},\ and\ \citenamefont {Thienpont}}]{szpulak2007single}%
  \BibitemOpen
  \bibfield  {author} {\bibinfo {author} {\bibfnamefont {M}~\bibnamefont
  {Szpulak}}, \bibinfo {author} {\bibfnamefont {T}~\bibnamefont {Martynkien}},
  \bibinfo {author} {\bibfnamefont {J}~\bibnamefont {Olszewski}}, \bibinfo
  {author} {\bibfnamefont {W}~\bibnamefont {Urbanczyk}}, \bibinfo {author}
  {\bibfnamefont {T}~\bibnamefont {Nasilowski}}, \bibinfo {author}
  {\bibfnamefont {Francis}\ \bibnamefont {Berghmans}}, \ and\ \bibinfo {author}
  {\bibfnamefont {H}~\bibnamefont {Thienpont}},\ }\bibfield  {title} {\enquote
  {\bibinfo {title} {Single-polarization single-mode photonic band gap
  fiber},}\ }\href@noop {} {\bibfield  {journal} {\bibinfo  {journal} {ACTA
  PHYSICA POLONICA SERIES A}\ }\textbf {\bibinfo {volume} {111}},\ \bibinfo
  {pages} {239} (\bibinfo {year} {2007})}\BibitemShut {NoStop}%
\bibitem [{\citenamefont {Chiles}\ and\ \citenamefont
  {Fathpour}(2016)}]{chiles2016demonstration}%
  \BibitemOpen
  \bibfield  {author} {\bibinfo {author} {\bibfnamefont {Jeff}\ \bibnamefont
  {Chiles}}\ and\ \bibinfo {author} {\bibfnamefont {Sasan}\ \bibnamefont
  {Fathpour}},\ }\bibfield  {title} {\enquote {\bibinfo {title} {Demonstration
  of ultra-broadband single-mode and single-polarization operation in
  t-guides},}\ }\href@noop {} {\bibfield  {journal} {\bibinfo  {journal}
  {Optics Letters}\ }\textbf {\bibinfo {volume} {41}},\ \bibinfo {pages}
  {3836--3839} (\bibinfo {year} {2016})}\BibitemShut {NoStop}%
\bibitem [{\citenamefont {Bassett}\ and\ \citenamefont
  {Argyros}(2002)}]{bassett2002elimination}%
  \BibitemOpen
  \bibfield  {author} {\bibinfo {author} {\bibfnamefont {Ian~M}\ \bibnamefont
  {Bassett}}\ and\ \bibinfo {author} {\bibfnamefont {Alexander}\ \bibnamefont
  {Argyros}},\ }\bibfield  {title} {\enquote {\bibinfo {title} {Elimination of
  polarization degeneracy in round waveguides},}\ }\href@noop {} {\bibfield
  {journal} {\bibinfo  {journal} {Optics Express}\ }\textbf {\bibinfo {volume}
  {10}},\ \bibinfo {pages} {1342--1346} (\bibinfo {year} {2002})}\BibitemShut
  {NoStop}%
\bibitem [{\citenamefont {Argyros}\ \emph {et~al.}(2004)\citenamefont
  {Argyros}, \citenamefont {Issa}, \citenamefont {Bassett},\ and\ \citenamefont
  {Van~Eijkelenborg}}]{argyros2004microstructured}%
  \BibitemOpen
  \bibfield  {author} {\bibinfo {author} {\bibfnamefont {A}~\bibnamefont
  {Argyros}}, \bibinfo {author} {\bibfnamefont {N}~\bibnamefont {Issa}},
  \bibinfo {author} {\bibfnamefont {I}~\bibnamefont {Bassett}}, \ and\ \bibinfo
  {author} {\bibfnamefont {MA}~\bibnamefont {Van~Eijkelenborg}},\ }\bibfield
  {title} {\enquote {\bibinfo {title} {Microstructured optical fiber for
  single-polarization air guidance},}\ }\href@noop {} {\bibfield  {journal}
  {\bibinfo  {journal} {Optics letters}\ }\textbf {\bibinfo {volume} {29}},\
  \bibinfo {pages} {20--22} (\bibinfo {year} {2004})}\BibitemShut {NoStop}%
\bibitem [{\citenamefont {Knight}(2003)}]{knight2003photonic}%
  \BibitemOpen
  \bibfield  {author} {\bibinfo {author} {\bibfnamefont {Jonathan~C}\
  \bibnamefont {Knight}},\ }\bibfield  {title} {\enquote {\bibinfo {title}
  {Photonic crystal fibres},}\ }\href@noop {} {\bibfield  {journal} {\bibinfo
  {journal} {nature}\ }\textbf {\bibinfo {volume} {424}},\ \bibinfo {pages}
  {847--851} (\bibinfo {year} {2003})}\BibitemShut {NoStop}%
\bibitem [{\citenamefont {Russell}(2006)}]{russell2006photonic}%
  \BibitemOpen
  \bibfield  {author} {\bibinfo {author} {\bibfnamefont {Philip St~J}\
  \bibnamefont {Russell}},\ }\bibfield  {title} {\enquote {\bibinfo {title}
  {Photonic-crystal fibers},}\ }\href@noop {} {\bibfield  {journal} {\bibinfo
  {journal} {Journal of lightwave technology}\ }\textbf {\bibinfo {volume}
  {24}},\ \bibinfo {pages} {4729--4749} (\bibinfo {year} {2006})}\BibitemShut
  {NoStop}%
\bibitem [{\citenamefont {Xie}\ \emph {et~al.}(2015)\citenamefont {Xie},
  \citenamefont {Zhang}, \citenamefont {Boardman}, \citenamefont {Jiang},
  \citenamefont {Hu}, \citenamefont {Liu}, \citenamefont {Xie}, \citenamefont
  {Mao}, \citenamefont {Hu}, \citenamefont {Li} \emph {et~al.}}]{xie2015fiber}%
  \BibitemOpen
  \bibfield  {author} {\bibinfo {author} {\bibfnamefont {Kang}\ \bibnamefont
  {Xie}}, \bibinfo {author} {\bibfnamefont {Wei}\ \bibnamefont {Zhang}},
  \bibinfo {author} {\bibfnamefont {Allan~D}\ \bibnamefont {Boardman}},
  \bibinfo {author} {\bibfnamefont {Haiming}\ \bibnamefont {Jiang}}, \bibinfo
  {author} {\bibfnamefont {Zhijia}\ \bibnamefont {Hu}}, \bibinfo {author}
  {\bibfnamefont {Yong}\ \bibnamefont {Liu}}, \bibinfo {author} {\bibfnamefont
  {Ming}\ \bibnamefont {Xie}}, \bibinfo {author} {\bibfnamefont {Qiuping}\
  \bibnamefont {Mao}}, \bibinfo {author} {\bibfnamefont {Lei}\ \bibnamefont
  {Hu}}, \bibinfo {author} {\bibfnamefont {Qian}\ \bibnamefont {Li}},  \emph
  {et~al.},\ }\bibfield  {title} {\enquote {\bibinfo {title} {Fiber guiding at
  the dirac frequency beyond photonic bandgaps},}\ }\href@noop {} {\bibfield
  {journal} {\bibinfo  {journal} {Light: Science \& Applications}\ }\textbf
  {\bibinfo {volume} {4}},\ \bibinfo {pages} {e304--e304} (\bibinfo {year}
  {2015})}\BibitemShut {NoStop}%
\bibitem [{\citenamefont {Biswas}\ \emph {et~al.}(2016)\citenamefont {Biswas},
  \citenamefont {Chattopadhyay},\ and\ \citenamefont
  {Bhadra}}]{biswas2016dirac}%
  \BibitemOpen
  \bibfield  {author} {\bibinfo {author} {\bibfnamefont {Tushar}\ \bibnamefont
  {Biswas}}, \bibinfo {author} {\bibfnamefont {Rik}\ \bibnamefont
  {Chattopadhyay}}, \ and\ \bibinfo {author} {\bibfnamefont {Shyamal~K}\
  \bibnamefont {Bhadra}},\ }\bibfield  {title} {\enquote {\bibinfo {title}
  {Dirac-mode guidance in silica-based hollow-core photonic crystal fiber with
  high-index dielectric rings},}\ }\href@noop {} {\bibfield  {journal}
  {\bibinfo  {journal} {physica status solidi (b)}\ }\textbf {\bibinfo {volume}
  {253}},\ \bibinfo {pages} {1898--1906} (\bibinfo {year} {2016})}\BibitemShut
  {NoStop}%
\bibitem [{\citenamefont {Lu}\ \emph {et~al.}(2013)\citenamefont {Lu},
  \citenamefont {Fu}, \citenamefont {Joannopoulos},\ and\ \citenamefont
  {Solja{\v{c}}i{\'c}}}]{lu2013weyl}%
  \BibitemOpen
  \bibfield  {author} {\bibinfo {author} {\bibfnamefont {Ling}\ \bibnamefont
  {Lu}}, \bibinfo {author} {\bibfnamefont {Liang}\ \bibnamefont {Fu}}, \bibinfo
  {author} {\bibfnamefont {John~D}\ \bibnamefont {Joannopoulos}}, \ and\
  \bibinfo {author} {\bibfnamefont {Marin}\ \bibnamefont
  {Solja{\v{c}}i{\'c}}},\ }\bibfield  {title} {\enquote {\bibinfo {title} {Weyl
  points and line nodes in gyroid photonic crystals},}\ }\href@noop {}
  {\bibfield  {journal} {\bibinfo  {journal} {Nature photonics}\ }\textbf
  {\bibinfo {volume} {7}},\ \bibinfo {pages} {294} (\bibinfo {year}
  {2013})}\BibitemShut {NoStop}%
\bibitem [{\citenamefont {Rechtsman}\ \emph {et~al.}(2013)\citenamefont
  {Rechtsman}, \citenamefont {Zeuner}, \citenamefont {Plotnik}, \citenamefont
  {Lumer}, \citenamefont {Podolsky}, \citenamefont {Dreisow}, \citenamefont
  {Nolte}, \citenamefont {Segev},\ and\ \citenamefont
  {Szameit}}]{rechtsman2013photonic}%
  \BibitemOpen
  \bibfield  {author} {\bibinfo {author} {\bibfnamefont {Mikael~C}\
  \bibnamefont {Rechtsman}}, \bibinfo {author} {\bibfnamefont {Julia~M}\
  \bibnamefont {Zeuner}}, \bibinfo {author} {\bibfnamefont {Yonatan}\
  \bibnamefont {Plotnik}}, \bibinfo {author} {\bibfnamefont {Yaakov}\
  \bibnamefont {Lumer}}, \bibinfo {author} {\bibfnamefont {Daniel}\
  \bibnamefont {Podolsky}}, \bibinfo {author} {\bibfnamefont {Felix}\
  \bibnamefont {Dreisow}}, \bibinfo {author} {\bibfnamefont {Stefan}\
  \bibnamefont {Nolte}}, \bibinfo {author} {\bibfnamefont {Mordechai}\
  \bibnamefont {Segev}}, \ and\ \bibinfo {author} {\bibfnamefont {Alexander}\
  \bibnamefont {Szameit}},\ }\bibfield  {title} {\enquote {\bibinfo {title}
  {Photonic floquet topological insulators},}\ }\href@noop {} {\bibfield
  {journal} {\bibinfo  {journal} {Nature}\ }\textbf {\bibinfo {volume} {496}},\
  \bibinfo {pages} {196--200} (\bibinfo {year} {2013})}\BibitemShut {NoStop}%
\bibitem [{\citenamefont {Noh}\ \emph {et~al.}(2017)\citenamefont {Noh},
  \citenamefont {Huang}, \citenamefont {Leykam}, \citenamefont {Chong},
  \citenamefont {Chen},\ and\ \citenamefont {Rechtsman}}]{noh2017experimental}%
  \BibitemOpen
  \bibfield  {author} {\bibinfo {author} {\bibfnamefont {Jiho}\ \bibnamefont
  {Noh}}, \bibinfo {author} {\bibfnamefont {Sheng}\ \bibnamefont {Huang}},
  \bibinfo {author} {\bibfnamefont {Daniel}\ \bibnamefont {Leykam}}, \bibinfo
  {author} {\bibfnamefont {Yi~Dong}\ \bibnamefont {Chong}}, \bibinfo {author}
  {\bibfnamefont {Kevin~P}\ \bibnamefont {Chen}}, \ and\ \bibinfo {author}
  {\bibfnamefont {Mikael~C}\ \bibnamefont {Rechtsman}},\ }\bibfield  {title}
  {\enquote {\bibinfo {title} {Experimental observation of optical weyl points
  and fermi arc-like surface states},}\ }\href@noop {} {\bibfield  {journal}
  {\bibinfo  {journal} {Nature Physics}\ }\textbf {\bibinfo {volume} {13}},\
  \bibinfo {pages} {611--617} (\bibinfo {year} {2017})}\BibitemShut {NoStop}%
\bibitem [{\citenamefont {Wu}\ and\ \citenamefont {Hu}(2015)}]{wu2015scheme}%
  \BibitemOpen
  \bibfield  {author} {\bibinfo {author} {\bibfnamefont {Long-Hua}\
  \bibnamefont {Wu}}\ and\ \bibinfo {author} {\bibfnamefont {Xiao}\
  \bibnamefont {Hu}},\ }\bibfield  {title} {\enquote {\bibinfo {title} {Scheme
  for achieving a topological photonic crystal by using dielectric material},}\
  }\href@noop {} {\bibfield  {journal} {\bibinfo  {journal} {Physical Review
  Letters}\ }\textbf {\bibinfo {volume} {114}},\ \bibinfo {pages} {223901}
  (\bibinfo {year} {2015})}\BibitemShut {NoStop}%
\bibitem [{\citenamefont {Joannopoulos}\ \emph {et~al.}(2008)\citenamefont
  {Joannopoulos}, \citenamefont {Johnson}, \citenamefont {Winn},\ and\
  \citenamefont {Meade}}]{joannopoulos2008molding}%
  \BibitemOpen
  \bibfield  {author} {\bibinfo {author} {\bibfnamefont {John~D}\ \bibnamefont
  {Joannopoulos}}, \bibinfo {author} {\bibfnamefont {Steven~G}\ \bibnamefont
  {Johnson}}, \bibinfo {author} {\bibfnamefont {Joshua~N}\ \bibnamefont
  {Winn}}, \ and\ \bibinfo {author} {\bibfnamefont {Robert~D}\ \bibnamefont
  {Meade}},\ }\bibfield  {title} {\enquote {\bibinfo {title} {Molding the flow
  of light},}\ }\href@noop {} {\bibfield  {journal} {\bibinfo  {journal}
  {Princeton Univ. Press, Princeton, NJ [ua]}\ ,\ \bibinfo {pages} {167}}
  (\bibinfo {year} {2008})}\BibitemShut {NoStop}%
\bibitem [{\citenamefont {Heiblum}\ and\ \citenamefont
  {Harris}(1975)}]{heiblum1975analysis}%
  \BibitemOpen
  \bibfield  {author} {\bibinfo {author} {\bibfnamefont {Mordehai}\
  \bibnamefont {Heiblum}}\ and\ \bibinfo {author} {\bibfnamefont {Jay}\
  \bibnamefont {Harris}},\ }\bibfield  {title} {\enquote {\bibinfo {title}
  {Analysis of curved optical waveguides by conformal transformation},}\
  }\href@noop {} {\bibfield  {journal} {\bibinfo  {journal} {IEEE Journal of
  Quantum Electronics}\ }\textbf {\bibinfo {volume} {11}},\ \bibinfo {pages}
  {75--83} (\bibinfo {year} {1975})}\BibitemShut {NoStop}%
\bibitem [{\citenamefont {Beravat}\ \emph {et~al.}(2016)\citenamefont
  {Beravat}, \citenamefont {Wong}, \citenamefont {Frosz}, \citenamefont {Xi},\
  and\ \citenamefont {Russell}}]{beravat2016twist}%
  \BibitemOpen
  \bibfield  {author} {\bibinfo {author} {\bibfnamefont {Ramin}\ \bibnamefont
  {Beravat}}, \bibinfo {author} {\bibfnamefont {Gordon~KL}\ \bibnamefont
  {Wong}}, \bibinfo {author} {\bibfnamefont {Michael~H}\ \bibnamefont {Frosz}},
  \bibinfo {author} {\bibfnamefont {Xiao~Ming}\ \bibnamefont {Xi}}, \ and\
  \bibinfo {author} {\bibfnamefont {Philip St~J}\ \bibnamefont {Russell}},\
  }\bibfield  {title} {\enquote {\bibinfo {title} {Twist-induced guidance in
  coreless photonic crystal fiber: A helical channel for light},}\ }\href@noop
  {} {\bibfield  {journal} {\bibinfo  {journal} {Science Advances}\ }\textbf
  {\bibinfo {volume} {2}},\ \bibinfo {pages} {e1601421} (\bibinfo {year}
  {2016})}\BibitemShut {NoStop}%
\bibitem [{\citenamefont {Finazzi}\ \emph {et~al.}(2003)\citenamefont
  {Finazzi}, \citenamefont {Monro},\ and\ \citenamefont
  {Richardson}}]{finazzi2003small}%
  \BibitemOpen
  \bibfield  {author} {\bibinfo {author} {\bibfnamefont {Vittoria}\
  \bibnamefont {Finazzi}}, \bibinfo {author} {\bibfnamefont {Tanya~M}\
  \bibnamefont {Monro}}, \ and\ \bibinfo {author} {\bibfnamefont {David~J}\
  \bibnamefont {Richardson}},\ }\bibfield  {title} {\enquote {\bibinfo {title}
  {Small-core silica holey fibers: nonlinearity and confinement loss
  trade-offs},}\ }\href@noop {} {\bibfield  {journal} {\bibinfo  {journal}
  {JOSA B}\ }\textbf {\bibinfo {volume} {20}},\ \bibinfo {pages} {1427--1436}
  (\bibinfo {year} {2003})}\BibitemShut {NoStop}%
\bibitem [{\citenamefont {Tsuchida}\ \emph {et~al.}(2005)\citenamefont
  {Tsuchida}, \citenamefont {Saitoh},\ and\ \citenamefont
  {Koshiba}}]{tsuchida2005design}%
  \BibitemOpen
  \bibfield  {author} {\bibinfo {author} {\bibfnamefont {Yukihiro}\
  \bibnamefont {Tsuchida}}, \bibinfo {author} {\bibfnamefont {Kunimasa}\
  \bibnamefont {Saitoh}}, \ and\ \bibinfo {author} {\bibfnamefont {Masanori}\
  \bibnamefont {Koshiba}},\ }\bibfield  {title} {\enquote {\bibinfo {title}
  {Design and characterization of single-mode holey fibers with low bending
  losses},}\ }\href@noop {} {\bibfield  {journal} {\bibinfo  {journal} {Optics
  express}\ }\textbf {\bibinfo {volume} {13}},\ \bibinfo {pages} {4770--4779}
  (\bibinfo {year} {2005})}\BibitemShut {NoStop}%
\bibitem [{\citenamefont {Teo}\ and\ \citenamefont
  {Kane}(2010)}]{teo2010topological}%
  \BibitemOpen
  \bibfield  {author} {\bibinfo {author} {\bibfnamefont {Jeffrey~CY}\
  \bibnamefont {Teo}}\ and\ \bibinfo {author} {\bibfnamefont {Charles~L}\
  \bibnamefont {Kane}},\ }\bibfield  {title} {\enquote {\bibinfo {title}
  {Topological defects and gapless modes in insulators and superconductors},}\
  }\href@noop {} {\bibfield  {journal} {\bibinfo  {journal} {Physical Review
  B}\ }\textbf {\bibinfo {volume} {82}},\ \bibinfo {pages} {115120} (\bibinfo
  {year} {2010})}\BibitemShut {NoStop}%
\bibitem [{\citenamefont {Yuan}\ \emph {et~al.}(2018)\citenamefont {Yuan},
  \citenamefont {Lin}, \citenamefont {Xiao},\ and\ \citenamefont
  {Fan}}]{yuan2018synthetic}%
  \BibitemOpen
  \bibfield  {author} {\bibinfo {author} {\bibfnamefont {Luqi}\ \bibnamefont
  {Yuan}}, \bibinfo {author} {\bibfnamefont {Qian}\ \bibnamefont {Lin}},
  \bibinfo {author} {\bibfnamefont {Meng}\ \bibnamefont {Xiao}}, \ and\
  \bibinfo {author} {\bibfnamefont {Shanhui}\ \bibnamefont {Fan}},\ }\bibfield
  {title} {\enquote {\bibinfo {title} {Synthetic dimension in photonics},}\
  }\href@noop {} {\bibfield  {journal} {\bibinfo  {journal} {Optica}\ }\textbf
  {\bibinfo {volume} {5}},\ \bibinfo {pages} {1396--1405} (\bibinfo {year}
  {2018})}\BibitemShut {NoStop}%
\end{thebibliography}%

\newpage


\section*{Supplementary material}

# Supplementary Information for "Dirac-vortex topological photonic crystal fibre"

Hao Lin[1,2] and Ling Lu[1,3,*]

[1]*Institute of Physics, Chinese Academy of Sciences/Beijing National Laboratory for Condensed Matter Physics, Beijing 100190*

[2]*School of Physical Sciences, University of Chinese Academy of Sciences, Beijing 100049, China*

[3]*Songshan Lake Materials Laboratory, Dongguan, Guangdong 523808, China*

## Abstract

**CONTENTS**

_____
* linglu@iphy.ac.cn

# I. THE SIGN OF WINDING NUMBER

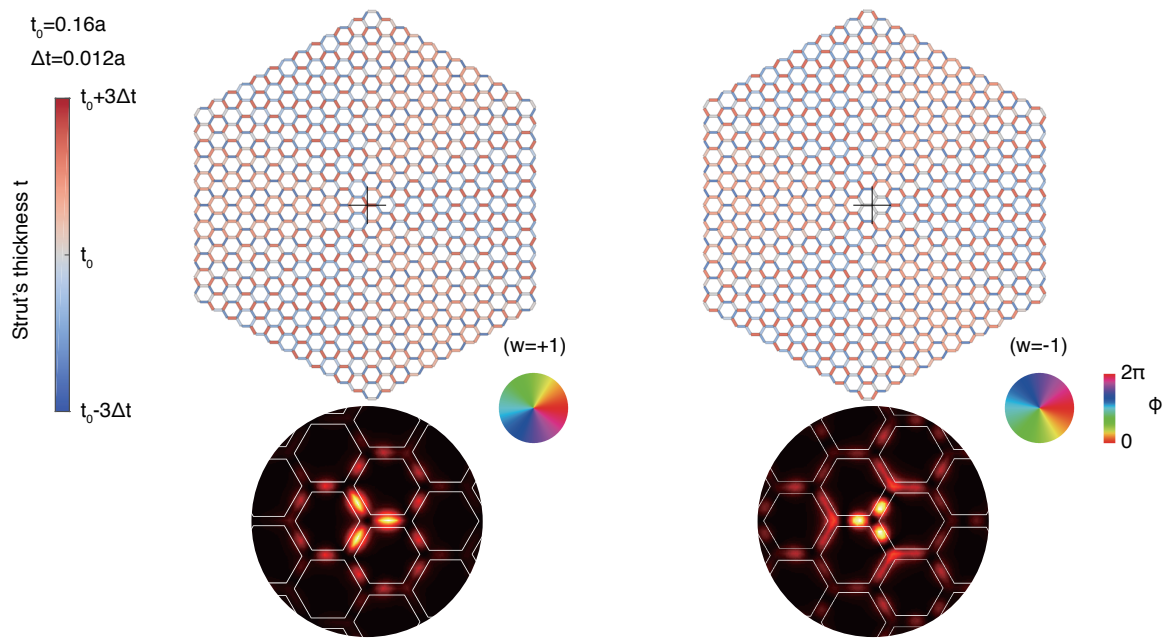

FIG. S1. Comparison of the field distributions for opposite winding numbers of the Dirac-vortex PCF. The intensity patterns localize on different strut joints, which are the two sub-lattices in a honeycomb lattice.

## II. DISCRETE MODULATION USING FOUR STRUT THICKNESSES

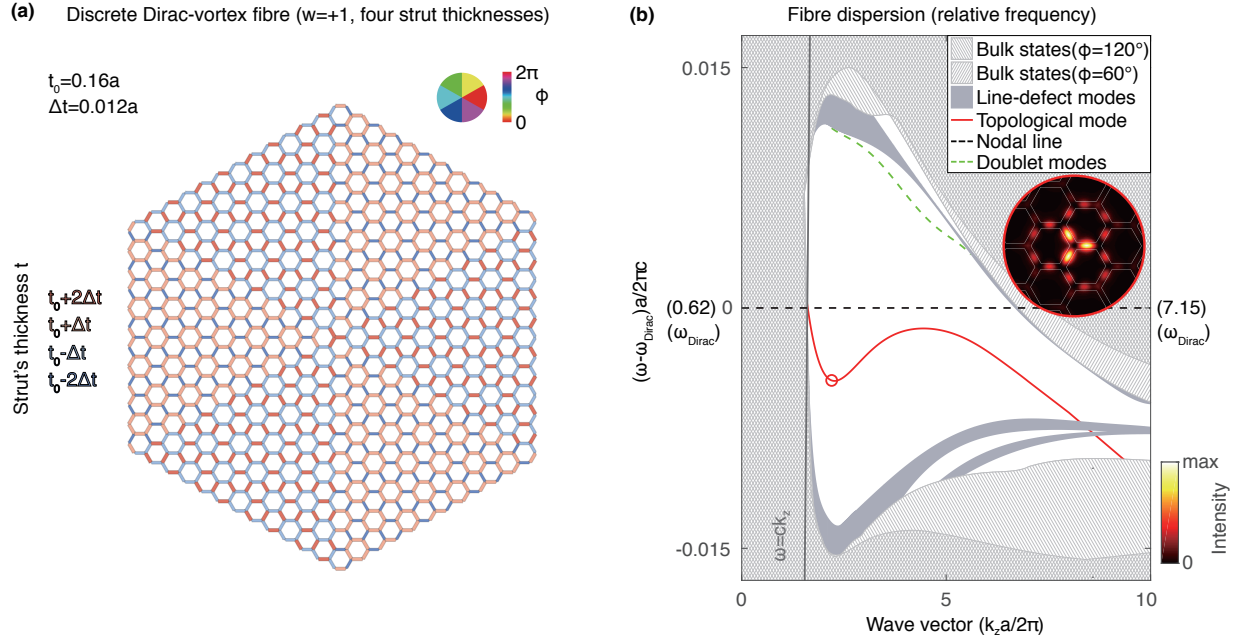

FIG. S2. Discrete Dirac-vortex fibre constructed by four strut thicknesses, instead of four tubes in Fig. 5 in the main-text. The cleaner band diagram is due to the lack of extra strut thicknesses from the tube construction of the stack-and-draw technique.

## III. HIGH-FREQUENCY DIRAC POINT

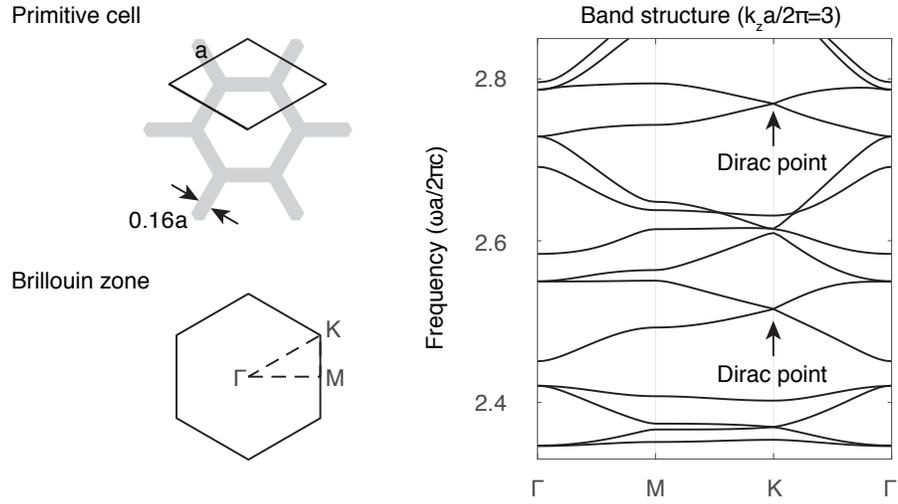

FIG. S3. Band structure of the honeycomb primitive cell at $k_z a/2\pi = 3$. The two frequency-isolated Dirac points correspond to the two topological dispersions in Fig. 6(b) in main text.

## IV. DESIGN TOLERANCE

Unlike the other topological waveguides, the Dirac-vortex fibre does not support one-way propagation nor sharp corner turning. The stability of the Dirac-vortex PCF lies in the design itself. First of all, this is essentially a coreless fibre [1] where the defect is not created by adding nor removing materials locally. The topological defect of the vortex is formed by gently perturbing the whole lattice globally, so that small local fabrication imperfection cannot unwind the vortex nor the vortex mode. Secondly, it is well known that the fibre drawing process will smoothen all the sharp edges, which we have not considered for the $120°$ corners between the neighboring two struts in the design. If we allow the rounded corners in the modeling [2, 3], the thickness difference between the struts decreases, so does the modulation amplitude responsible for the gap opening. Consequently, although the topological gap shrinks as the curvature radius increases, it remains open for typical curvature values, as show in Fig. S4(a). Structure in Fig. S4(b) is a discrete vortex fibre (from four-tube construction) with curvature radius $r' = 0.3a$, obtained by adding extra material to the design in Fig. 5(b) in the main-text. The corresponding dispersion and loss performance are shown in Fig. S4(c) and Fig. S4(d).

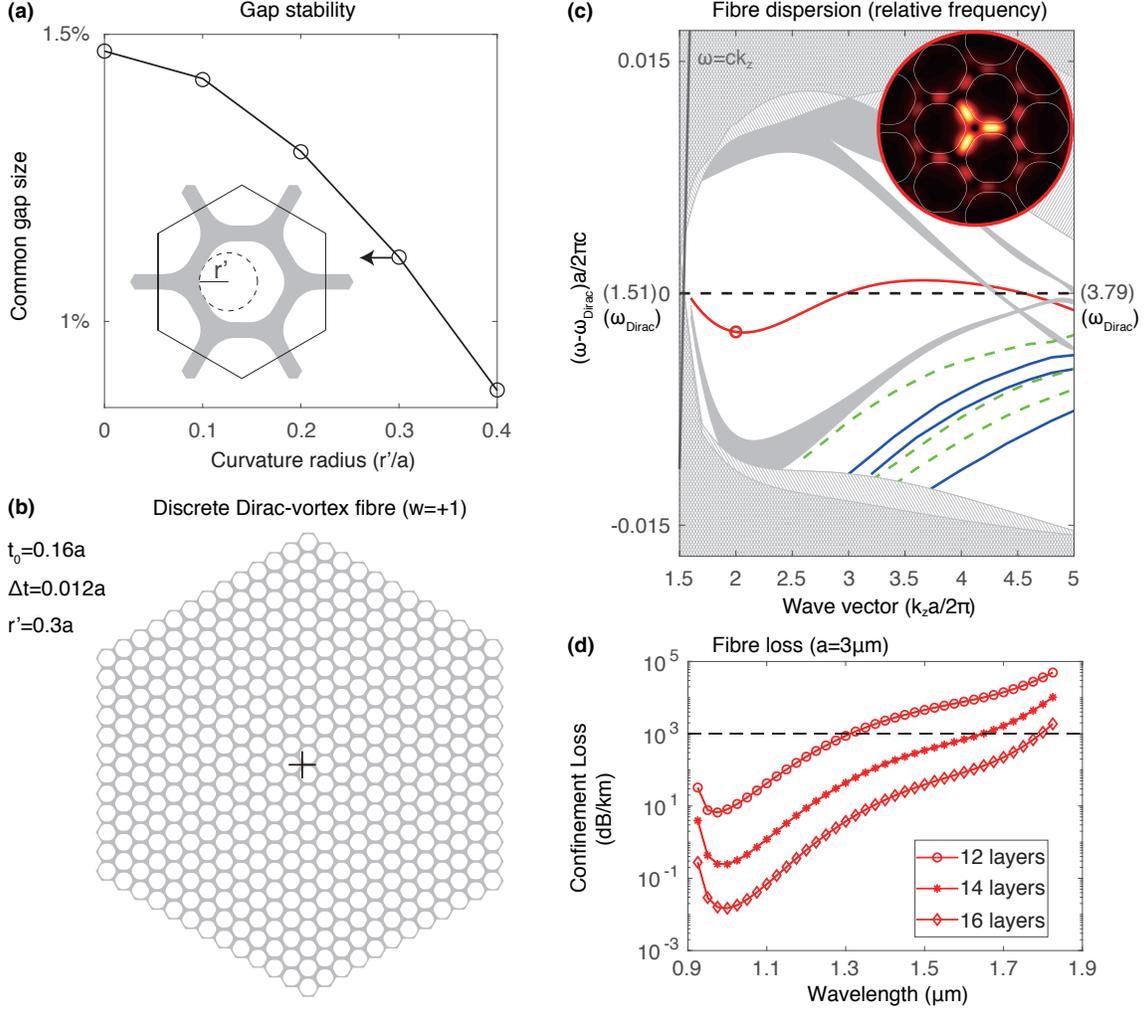

FIG. S4. Design tolerance of structural curvature ($r'$) common in fibre drawing process. (a) Size of common Dirac-vortex bandgap, for all $\phi$ at $k_z a/2\pi = 2$, as a function of the curvature radius of corners. Inset: supercell with $r' = 0.3a$ and $t_0 = 0.16a$. (b) Fibre structure with $r' = 0.3a$ by adding corner material to the discrete fibre in Fig. 5(b) in the main-text. (c) Band diagram with a frequency reference to the original nodal-line. The dispersion is similar to that in Fig. 5(c). Inset shows the mode profile ($\hat{z} \cdot \mathrm{Re}[\boldsymbol{E}^* \times \boldsymbol{H}]$). (d) Confinement loss of the topological mode. The wavelength range corresponds to the wave vector in (c) for $k_z a/2\pi$ from 1.6 to 4.2.

## V. OPERATION ABOVE LIGHT LINE

One key feature of PCFs is the ability to support bandgaps above the light line for hollow-core modes. The topological PCF can operate above the light line as well. In Fig. S5, we pick a particular set of geometric parameters ($t_0 = 0.12a$, $\Delta t = 0.01a$) to push the topological bandgap, and the topological guiding mode, into the light cone. We also calculate the modal concentration factor in air for the topological mode. The maximum air fraction is about 75% for this design, where the effective modal index $n_{\text{eff}} < 1$. In order to have more bandwidth above the light line, we need to improve the frequency isolation of the nodal line degeneracy for low frequencies, so the project bulk bands can remain gapped. This should be possible by increasing the refractive index of glass or explore a different lattice structure.

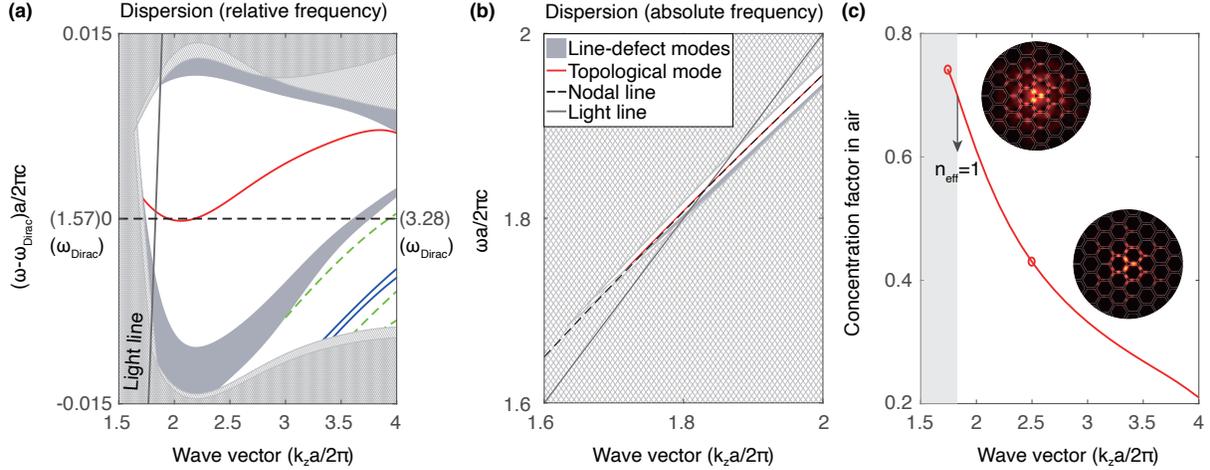

FIG. S5. Topological fibre mode above light line. (a) Band diagram of a discrete Dirac-vortex fibre with $t_0 = 0.12a$ and $\Delta t = 0.01a$ (the discrete version of four tubes). (b) Dispersion of absolute frequency near the light line (c) Concentration of the intensity of the mode in air. Insets: mode intensity at $k_z a / 2\pi = 1.74$ and 2.5.

[1] Ramin Beravat, Gordon KL Wong, Michael H Frosz, Xiao Ming Xi, and Philip St J Russell, "Twist-induced guidance in coreless photonic crystal fiber: A helical channel for light," Science Advances **2**, e1601421 (2016).

[2] Niels Asger Mortensen and Martin Dybendal Nielsen, "Modeling of realistic cladding structures for air-core photonic bandgap fibers," Optics letters **29**, 349–351 (2004).

[3] Francesco Poletti, Neil GR Broderick, David James Richardson, and Tanya Mary Monro, "The effect of core asymmetries on the polarization properties of hollow core photonic bandgap fibers," Optics Express **13**, 9115–9124 (2005).



\end{document}